\documentclass[letter,longauth]{aa} 
\usepackage{graphicx}
\usepackage{ulem}
\usepackage[varg]{txfonts}
\usepackage{natbib}
\usepackage{multirow}

\usepackage{enumerate}
\usepackage{booktabs}
\usepackage{lscape}

\newcommand{\kboltz}{k_{\rm B}}

\newcommand{\CO}{\ensuremath{^{12}\mathrm{CO}}}
\newcommand{\tCO}{\ensuremath{^{13}\mathrm{CO}}}
\newcommand{\CeO}{\ensuremath{\mathrm{C^{18}O}}}

\newcommand{\NtHp}{\ensuremath{\mathrm{N_{2}H^{+}}}}
\newcommand{\Tdust}{\ensuremath{T_{\rm dust}}}

\newcommand{\Tmb}{\ensuremath{ T_{\rm mb} }}
\newcommand{\Tex}{\ensuremath{ T_{\rm ex} }}

\newcommand{\Vlsr}{\ensuremath{ v_{\rm lsr} }}

\begin{document}

   \title{Multi-scale analysis of the Monoceros OB 1 star-forming region}
   \subtitle{I. The dense core population}
   \author{Julien Montillaud\inst{1}
         \and
         Mika Juvela\inst{2}
         \and
         Charlotte Vastel\inst{3}
         \and
         J.H. He\inst{4,5,6} 
         \and
         Tie Liu\inst{7,8,9}
         \and
         Isabelle Ristorcelli\inst{3}
         \and
         David J. Eden\inst{10}
         \and
         Sung-ju Kang\inst{8}
         \and
         Kee-Tae Kim\inst{8,11}
         \and
         Patrick M. Koch\inst{12}
         \and
         Chang Won Lee\inst{8,11}
         \and
         Mark G. Rawlings\inst{9}
         \and
         Mika Saajasto\inst{2}
         \and
         Patricio Sanhueza\inst{13}
         \and
         Archana Soam\inst{14,8}
         \and
         Sarolta Zahorecz\inst{15,13}
         \and
         Dana Alina\inst{16}
         \and
         Rebeka Bögner\inst{1,17}
         \and
         David Cornu\inst{1}
         \and
         Yasuo Doi\inst{18}
         \and
         Johanna Malinen\inst{19}
         \and
         Douglas Marshall\inst{20}
         \and
         E. R. Micelotta\inst{2}
         \and
         V.M. Pelkonen\inst{1,2,21}
         \and
         L. V. Tóth\inst{17,22}
         \and
         A. Traficante\inst{23}
         \and
         Ke Wang\inst{24,25}
         }


   \institute{
      Institut UTINAM - UMR 6213 - CNRS - Univ Bourgogne Franche Comté, France, OSU THETA, 41bis avenue de l'Observatoire, 25000 Besan\c{c}on, France - e-mail: {\tt julien@obs-besancon.fr}
      \and 
      Department of Physics, P.O.Box 64, FI-00014, University of Helsinki, Finland
      \and 
      IRAP, Université de Toulouse, CNRS, UPS, CNES, F-31400 Toulouse, France
      \and 
      Yunnan Observatories, Chinese Academy of Sciences, 396 Yangfangwang, Guandu District, Kunming, 650216, P. R. China
      \and 
      Chinese Academy of Sciences, South America Center for Astrophysics (CASSACA), Camino El Observatorio 1515, Las Condes, Santiago, Chile
      \and 
      Departamento de Astronom\'ia, Universidad de Chile, Las Condes, Santiago, Chile
      \and 
      Shanghai Astronomical Observatory, Chinese Academy of Sciences, 80 Nandan Road, Shanghai 200030, China
      \and 
      Korea Astronomy and Space Science Institute, 776 Daedeokdaero, Yuseong-gu, Daejeon 34055, Republic of Korea
      \and 
      East Asian Observatory, 660 N. A’ohoku Place, Hilo, HI 96720, USA
      \and 
      Astrophysics Research Institute, Liverpool John Moores University, IC2, Liverpool Science Park, 146 Brownlow Hill, Liverpool L3 5RF, UK
      \and 
      University of Science \& Technology, 176 Gajeong-dong, Yuseong-gu, Daejeon, Republic of Korea
      \and 
      Academia Sinica, Institute of Astronomy and Astrophysics, Taipei, Taiwan
      \and 
      National Astronomical Observatory of Japan, National Institutes of Natural Sciences, 2-21-1 Osawa, Mitaka, Tokyo 181-8588, Japan
      \and 
      SOFIA Science Centre, USRA, NASA Ames Research Centre, MS N232-12 Moffett Field, CA 94035, USA
      \and 
      Department of Physical Science, Graduate School of Science, Osaka Prefecture University, 1-1 Gakuen-cho, Naka-ku, Sakai, Osaka 599-8531, Japan
      \and 
      Department of Physics, School of Science and Humanities, Kabanbay batyr ave, 53, Nur-Sultan 010000 Kazakhstan
      \and 
      Eötvös Loránd University, Department of Astronomy, Pázmány Péter sétány 1/A, H-1117, Budapest, Hungary
      \and 
      Department of Earth Science and Astronomy, Graduate School of Arts and Sciences, The University of Tokyo, 3-8-1 Komaba, Meguro, Tokyo 153-8902, Japan
      \and 
      Institute of Physics I, University of Cologne, Zülpicher Str. 77, D-50937, Cologne, Germany
      \and 
      AIM, CEA, CNRS, Université Paris-Saclay, Université Paris Diderot, Sorbonne Paris Cité, F-91191 Gif-sur-Yvette, France
      \and 
      ICC, University of Barcelona, Marti i Franquès 1, E-08028 Barcelona, Spain
      \and 
      Konkoly Observatory of the Hungarian Academy of Sciences, H-1121 Budapest, Konkoly Thege Miklósút 15-17, Hungary
      \and 
      IAPS-INAF, via Fosso del Cavaliere 100, I-00133, Rome, Italy
      \and 
      Kavli Institute for Astronomy and Astrophysics, Peking University, 5 Yiheyuan Road, Haidian District, Beijing 100871, China
      \and 
      European Southern Observatory (ESO) Headquarters, Karl-Schwarzschild-Str. 2, 85748 Garching bei M\"{u}nchen, Germany
      }

\date{Received 25 July 2019; Accepted 9 September 2019}
   \abstract
   {  Current theories and models attempt to explain star formation globally,
   from core scales to giant molecular cloud scales. A multi-scale observational
   characterisation of an entire molecular complex is necessary to constrain them.
   We investigate star formation in G202.3+2.5, a $\sim 10 \times 3\,\rm pc$ 
   sub-region of the Monoceros OB1 cloud with a complex morphology harbouring
   interconnected filamentary structures. }
   { We aim to connect the evolution of cores and filaments in G202.3+2.5 with
   the global evolution of the cloud and to identify the engines of the cloud
   dynamics.}
   { In this first paper, the star formation
   activity is evaluated by surveying the distributions of dense cores and 
   protostars, and their evolutionary state, as characterised using both 
   infrared observations from the {\it Herschel} and WISE telescopes and molecular 
   line observations with the IRAM 30-m telescope.
   }
   { We find ongoing star formation in the whole cloud, with a local peak in star
   formation activity around the centre of G202.3+2.5 where a chain of massive
   cores ($10-50$\,M$_\odot$) forms a massive ridge ($\gtrsim 150$\,M$_\odot$).
   All evolutionary stages from starless cores to Class II protostars are found
   in G202.3+2.5, including a possibly starless, large column density ($8 \times 10^{22}$
   cm$^{-2}$), and massive (52 M$_\odot$) core.
   }
   { All the core-scale observables examined in this paper point to an enhanced
   star formation activity centred on the junction between the three
   main branches of the ramified structure of G202.3+2.5. This suggests that the 
   increased star-formation activity results from the convergence of these branches. 
   To further investigate the origin of this enhancement, it is now necessary to 
   extend the analysis to larger scales, in order to examine the relationship 
   between cores, filaments and their environment. We address these points through 
   the analysis of the dynamics of G202.3+2.5 in a joint paper.
   }


   \maketitle
%


\section{Introduction}  \label{sec:introduction} 

   Star formation is a key topic in astrophysics that has stimulated the development
   of many observational programmes and simulation codes. The dramatic increase
   in computing capabilities over the last decades has made large-scale and 
   high-resolution simulations tractable \citep[e.g.][]{vazquez-semadeni_high-_2009,
   vazquez-semadeni_global_2019, renaud_sub-parsec_2013, padoan_supernova_2017}, in contrast to
   observational projects generally more focused on a given scale. Although 
   demanding in observing time, multi-scale observational programmes are now
   necessary for proper comparison with models. The feasibility and 
   fruitfulness of such multi-scale observational programmes was demonstrated 
   for example by the ORION B project \citep{pety_anatomy_2017,orkisz_turbulence_2017,
   orkisz_dynamically_2019, gratier_dissecting_2017} with the IRAM-30m telescope,
   the GAS survey with the GBT 100-m telescope on Gould Belt clouds \citep
   {friesen_green_2017}, or the high-resolution study of the infrared dark cloud
   SDC13 with the JVLA interferometer \citep{williams_evolution_2018}. Increasing
   the number and diversity of studied star formation regions is a key to improve
   our understanding of star formation.

   We took advantage of the Planck Galactic cold clump catalogue \citep[PGCC, ][]
   {montier_all-sky_2010, planck_collaboration_planck_2011, planck_collaboration_planck_2016} 
   and its {\it Herschel} follow-up in the frame of the open time key programme
   Galactic cold cores \citep[GCC, ][]{juvela_galactic_2012} to select G202.3+2.5, 
   a nearby filament \citep[760 pc, ][]{montillaud_galactic_2015} at the edge of 
   the Monoceros OB1 molecular complex. Its morphology is suggestive of a complex 
   dynamics (Fig.~\ref{fig:source_distrib}), and an intense star formation activity 
   was revealed there by the identification of almost a hundred candidate cold
   dense cores \citep{montillaud_galactic_2015}. Its rich environment includes
   the young ($\sim 3$ Myr) open cluster NGC 2264, an HII region and a large
   reservoir of molecular gas. A multi-scale study of this region is needed
   to shed light on the interplay between the cores, the filaments, and their
   environment.

   In this paper, the first of a series of papers on the Monoceros OB 1 region,
   we focus on the core scale ($\sim 0.1$ pc) and characterise the dense 
   cores in the G202.3+2.5 filaments. We analyse the dust emission in the far-IR 
   observed by {\it Herschel} and in the mid-IR by WISE over a 0.5 deg$^2$ area 
   covering a physical length of approximately 10 pc along the filament. This is 
   combined with the molecular gas emission observed in the millimetre range 
   with the IRAM 30-m telescope. The dynamics at larger scales and its relationship
   with the core scale are addressed in the next papers of the series.

\section{Observations}\label{sec:observations} 

   G202.3+2.5 was mapped with the {\it Herschel} instruments SPIRE
   \citep[250\,$\mu$m, 350\,$\mu$m and 500\,$\mu$m, ]{griffin_herschel-spire_2010} 
   and PACS \citep[100\,$\mu$m and 160\,$\mu$m, ]{poglitsch_photodetector_2010}, 
   as part of the \textit{Herschel} open time key programme Galactic cold cores 
   \citep{juvela_galactic_2010}. The data were reduced as explained by \citet
   {juvela_galactic_2012}. The resolutions are 18, 25 and 37$\arcsec$ (0.07, 
   0.09, 0.14 pc) for the 250, 350 and 500\,$\mu$m bands of SPIRE, and
   of 7.7 and 12$\arcsec$ (0.03, 0.04 pc) for the 100 and 160\,$\mu$m 
   bands of PACS. The calibration accuracies of the {\it Herschel}
   SPIRE and PACS surface brightness are expected to be better than 7\%\footnote
   {http://herschel.esac.esa.int/Documentation.shtml} and 10\%\footnote{
   http://Herschel.esac.esa.int/twiki/bin/view/Public/PacsCalibrationWeb},
   respectively.

   Part of the G202.3+2.5 cloud was observed with the IRAM 30-m telescope during 
   March 2017 at 110 GHz with the EMIR receiver, targeting the lines of 
   \tCO\,(J=1-0) and \CeO\,(J=1-0). EMIR was connected to both the VESPA 
   and FTS autocorrelators with spectral resolutions $\Delta \nu=20$ and 200 kHz 
   (0.055 km\,s$^{-1}$ and 0.55 km\,s$^{-1}$ at 110 GHz), respectively. The wide passband of the
   FTS enabled us to also observe $^{12}$CO (J=1-0) and \NtHp (J=1-0) lines. 
   Table~\ref{tab:lines} summarises the observations.

   We observed 14 tiles of typically $200\arcsec \times 180 \arcsec$, and built 
   a mosaic which covers some $130$ arcmin$^2$ around
   the position $(\alpha,\delta)_{J2000} = (6^h41^m00^s.5, +10^{\circ}42\arcmin
   27\arcsec)$. Each tile was observed multiple times and in orthogonal directions
   in on-the-fly (OTF) mode and position switching mode, with a scan velocity of
   at most 9\arcsec/s, a dump time of 1s and a maximum row spacing of 12\arcsec.
   The beam full width at half maximum (FWHM) ranges from $21\arcsec$ (0.08 pc) at 
   115 GHz to $26\arcsec$ (0.1 pc) at 93 GHz.
   The off position was observed every 1 to 1.5 minutes. It was chosen at 
   $(\alpha,\delta)_{J2000} = (6^h42^m30^s.36, +10^{\circ}33\arcmin 21.2\arcsec)$, 
   after searching the SPIRE 250\,$\mu$m map for a minimum in surface brightness. 
   Pointing corrections and focus corrections were performed
   every 1.5h and 3h, respectively, leading to a pointing accuracy measured to
   be $\lesssim 5\arcsec$. We converted the antenna temperature to main beam 
   temperature assuming a standard telescope main beam efficiency\footnote 
   {see http://www.iram.es/IRAMES/mainWiki/Iram30mEfficiencies} of 0.78 for CO
   observations and 0.80 for \NtHp.

   We also used archival data from the Wide-Field Infrared Survey Explorer 
   (WISE) satellite \citep{wright_wide-field_2010} at 3.4, 4.6, 12, and 22\,$\mu$m 
   with spatial resolution ranging from 6.1\arcsec (0.02 pc) at the shortest 
   wavelength to 12\arcsec (0.04 pc) at 22\,$\mu$m. We use these data to 
   complement the spectral energy distributions (SEDs) of cores in
   the mid-IR range. The data were converted to surface brightness units with 
   the conversion factors given in the explanatory supplement \citep
   {cutri_explanatory_2011}. The calibration uncertainty is approximately 6\% for the 
   22\,$\mu$m band and less for the shorter wavelengths.

   \begin{center}
   \begin{table}
      \caption{List of detected lines in our IRAM observations and used in this paper. }
      \begin{tabular}{c | c | c | c | c | c}
         \toprule
                           &  $\nu$    & Backend      & $\Delta v$      &  \Tmb$^{\rm max}$   & rms    \\
                           &   (GHz)   &              & (km s$^{-1}$)   &   (K)               & (mK)   \\
         \midrule
         \CO               &  115.271  & FTS          &  0.52           &  25.3          &  50 - 100    \\
         \tCO              &  110.201  & FTS          &  0.54           &  10.8          &  50 - 100    \\
                           &           &     VESPA    &       0.054     &  11.9          &  150 - 300   \\
         \CeO              &  109.782  & FTS          &  0.54           &   2.6          &  50 - 100    \\
                           &           &     VESPA    &       0.054     &   4.1          &  150 - 300   \\
         \NtHp             &  93.1763  & FTS          &  0.64           &   1.8          &  50 - 100    \\
         \bottomrule
      \end{tabular}
      \tablefoot{ 
      The beam FWHM ranges from $21\arcsec$ at 115 GHz to $26\arcsec$ at 93 GHz.
      The columns are:
      (1) Name of the species. All the transitions are J=1-0. 
      (2) Frequency of the transition, from the CDMS database \citep{endres_cologne_2016},
      except for \NtHp, from \citet{pagani_frequency_2009}. For transitions with
      multiple components, $\nu$ is given for the component with the highest
      frequency.
      (3) Spectrometer used to record the data. FTS and VESPA were used at 
      resolutions of 200 kHz and 20 kHz, respectively.
      (4) Velocity resolution. All spectra were resampled with channels of 
      0.6 km\,s$^{-1}$ (FTS) or 0.06 km\,s$^{-1}$ (VESPA).
      (5) Maximum main beam temperature of the transition in the map. 
      (6) Approximate rms range computed from a 10 km\,s$^{-1}$ range, at least 20 km\,s$^{-1}$
      off the line, where no astronomical signal is found.
      }
      \label{tab:lines}
   \end{table}
   \end{center}

\vspace{-1cm}
\section{Method}\label{sec:method} 

\subsection{Analysis of the molecular line data} 

   For the following analysis, we used multi-component Gaussian fits to 
   characterise the emission lines. For compact sources, spectra in all CO 
   isotopologues were first averaged within the ellipse of the source as defined
   in the GCC catalogue by \citet{montillaud_galactic_2015}. A combined, 
   single-component Gaussian fit of the \CO, \tCO\, and \CeO\, lines with the same
   centroid for the three lines was attempted. If the residuals show features 
   with signal-to-noise ratio (SNR) greater than five, a new fit with one more 
   component was done, again using the same
   centroid for all three lines. The residuals were examined again, and new fits
   with additional components were attempted until the residuals were below 
   5$\sigma$. We found a maximum of three components. Examples for a few
   sources are shown in Sect.~\ref{sec:fit_examples}. The 
   constraint that each Gaussian component must have the same centroid in all
   CO isotopologues was motivated by the following virial analysis which combines
   characteristics of two isotopologues. Although the \CeO\,emission is generally
   weak, it was found to help significantly the fitting procedure to identify
   reasonable components, especially when two velocity components were not well
   separated in the spectra of the other isotopologues.  
   Finally, to avoid including extended structures, only 
   components with a significantly peaked radial profile were included 
   (Sect.~\ref{sec:fit_examples}).

   Following the method described by \citet{wilson_tools_2013}, we assumed
   that all CO isotopologues share the same excitation temperature \Tex~which
   was derived from the (assumed) optically thick \CO\,(1-0) line. The optical
   depth of the \tCO\, and \CeO\,(1-0) lines were then obtained, and their column
   densities were derived (Sect.~\ref{sec:TexColdens}). The line widths were used
   as proposed by \citet{maclaren_corrections_1988} to derive the virial
   parameters by separating the thermal and non-thermal velocity dispersions
   (Sect.~\ref{sec:virial_analysis}).

\subsection{Submillimetre compact sources}\label{sec:GCCsources} 

   To characterise the star formation activity of G202.3+2.5, we investigated
   the distribution of dense cores and young stellar objects (YSOs) in the cloud, 
   and their stages of evolution (starless or protostellar). We do not
   intend to provide an accurate classification  (Class I, Class II, \dots)
   of each embedded YSO, but we use such classification from available public 
   catalogues.

   \citet{montillaud_galactic_2015} compiled the GCC catalogue which contains
   4466 compact sources extracted from the GCC \textit{Herschel} maps, including
   that of G202.3+2.5. They provide the physical characteristics of each
   source. It includes the mass, dust temperature (\Tdust), and the evolutionary
   stage. Because of the large number of sources and of the variety in 
   their distances, the classification criteria provide good statistical 
   results, but could be optimised for individual clouds.

   In the present study we focus on G202.3+2.5 where 98 sources were listed
   in the GCC catalogue, making it possible to examine more carefully individual
   sources. In Sect.~\ref{sec:GCCsources_method}, we present a refinement of the
   classification method adopted by \citet{montillaud_galactic_2015}, 
   based on the analysis of the source SED from near-IR (NIR) to far-IR (FIR)
   wavelengths. We define starless cores as dense sources without protostars, 
   regardless of their ability to eventually form a star, and protostellar cores
   as dense sources hosting at least one YSO. We also disentangle between
   candidate protostellar (or starless) sources, for which the proposed 
   evolutionary stage still remains to be confirmed, and reliable protostellar 
   (or starless) sources, when several independent diagnoses lead to the same 
   conclusion (Sect.~\ref{sec:GCCsources_method}).


\section{Results}\label{sec:results} 

   \label{sec:results_source_class}

   \begin{figure}[h!]
      \includegraphics[width=0.5\textwidth]{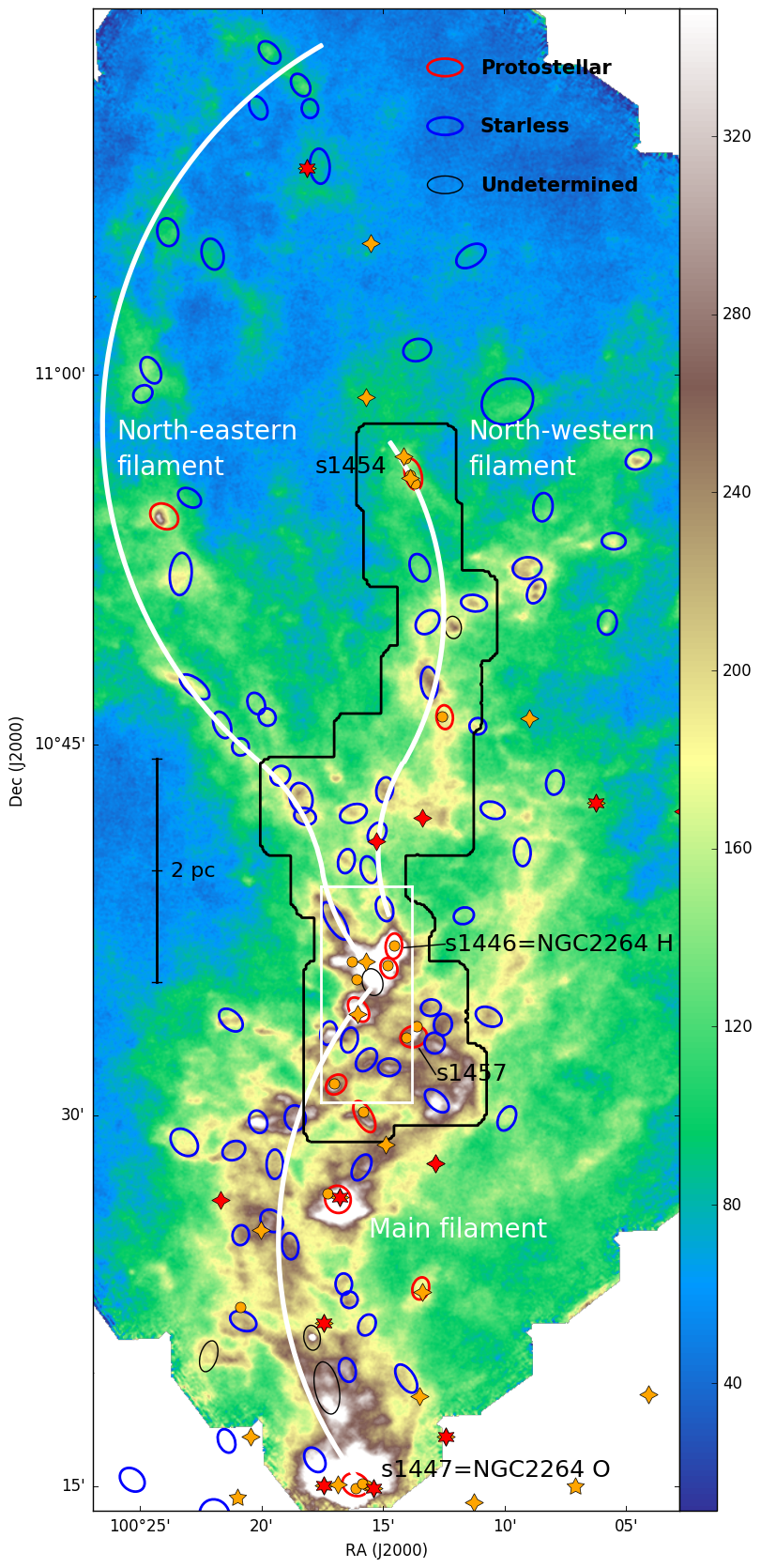}
      \caption{Compact and point source distribution in G202.3+2.5 overlaid to the
      surface brightness map of PACS 160\,$\mu$m in MJy/sr. Red, blue, and black 
      ellipses show GCC sources classified as protostellar, starless, and 
      undetermined. Red four- and six-pointed stars correspond, respectively, 
      to Class I/II and Class III (or more) YSOs from \citet{marton_all-sky_2016}. 
      Orange symbols are from \citet{rapson_spitzer_2014} and circles, four-, 
      and five-pointed stars correspond to Class 0/I, Class II, and transition disks, 
      respectively. The black contour shows the coverage of the IRAM 
      observations. The curved white lines sketch the general structure of 
      the cloud. The white rectangle shows the junction region as represented
      in Fig.~\ref{fig:source_class_example}.
      }
      \label{fig:source_distrib}
   \end{figure}

   \begin{figure*}[h!]
      \includegraphics[height=0.60\textheight]{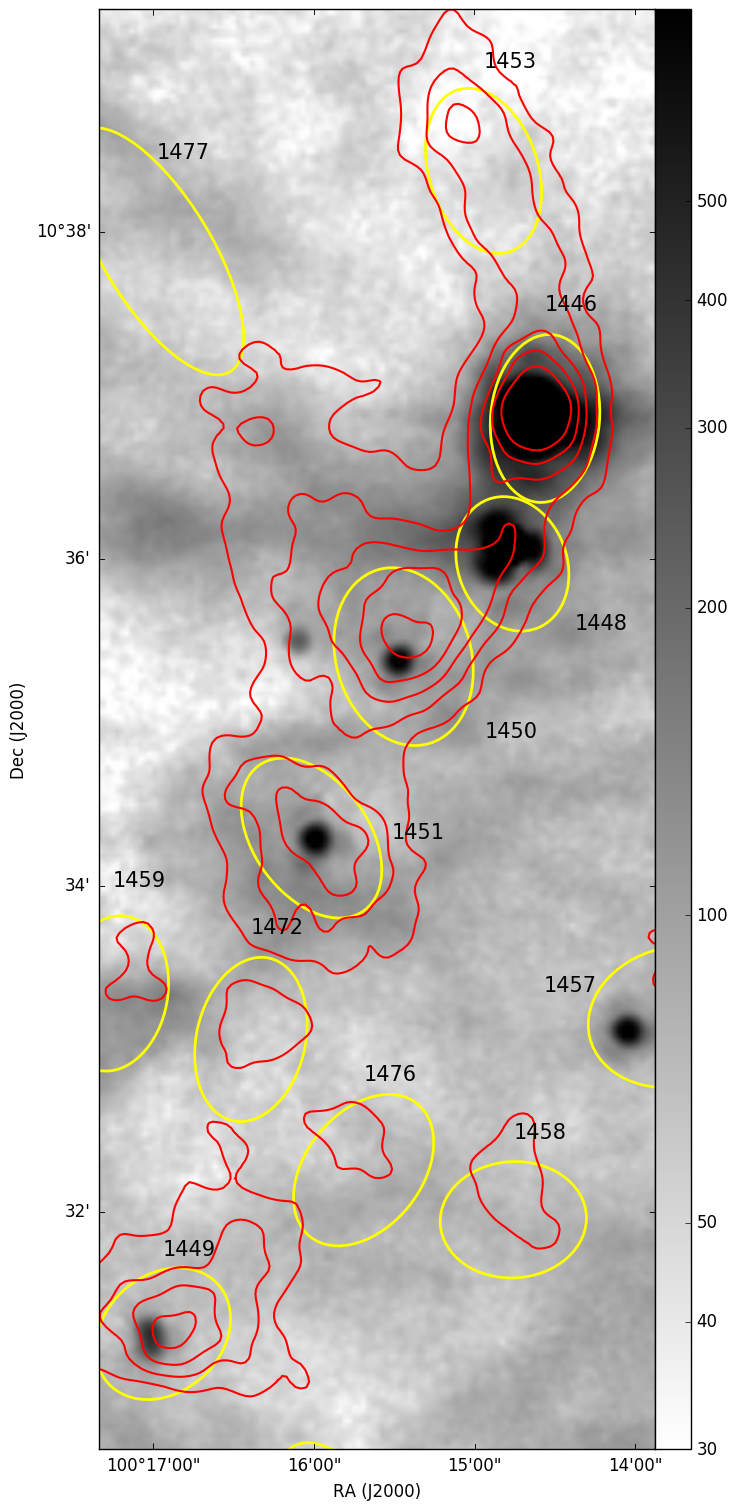}
      \hfill
      \includegraphics[height=0.60\textheight]{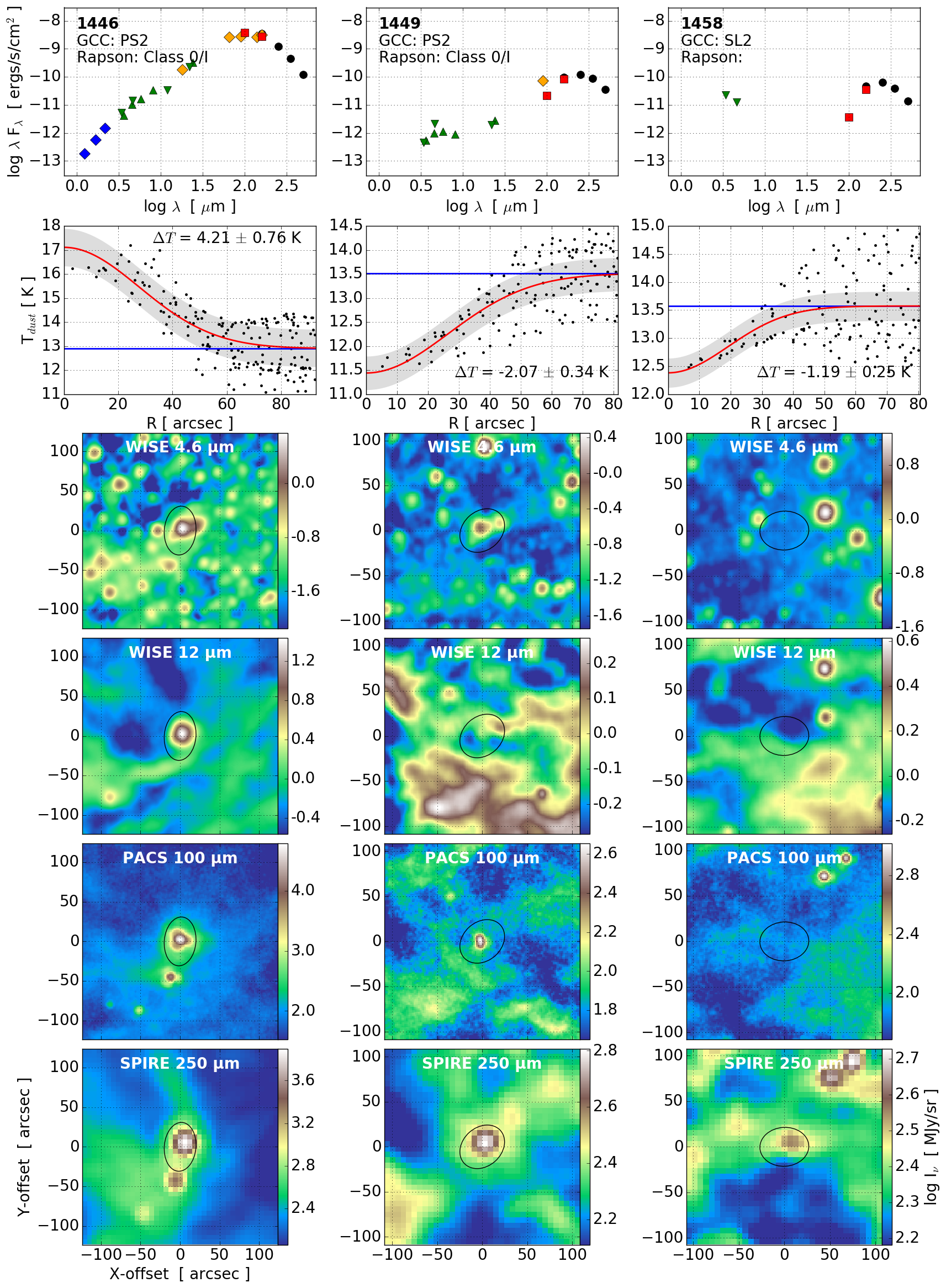}
      \caption{\textit{Left:} Surface brightness map (in MJy/sr) of PACS 100\,$\mu$m 
      in the junction region. The colour scale is cut to reveal 
      the cloud structure since the brightest source (s1446) peaks over 60000 
      MJy/sr. The red contours show the integrated emission of \NtHp\, at 1, 2, 3, 4,
      and 5 K\,km\,s$^{-1}$. The yellow ellipses show the FWHM of the GCC sources, 
      whose ID numbers are written in black. {\it Right:} The three columns show,
      for sources 1446, 1449 and 1458: the source SED, its \Tdust\, profile, 
      and maps of WISE 3.4\,$\mu$m, WISE 12\,$\mu$m, PACS 100\,$\mu$m and 
      SPIRE 250\,$\mu$m bands. In the SEDs: Blue diamonds are 2MASS J, H, and Ks 
      fluxes from the 2MASS PSC. Green tip-up triangles are {\it Spitzer}
      3.6, 4.5, 5.8, 8.0, and 24\,$\mu$m fluxes from the YSO catalogue of \citet
      {rapson_spitzer_2014}. Green tip-down triangles are WISE 3.4, 4.6, 12, and 22
      \,$\mu$m fluxes from our photometry measurements. Orange diamonds are AKARI
      9, 18, 65, 90, 140, and 160\,$\mu$m fluxes from the AKARI MIR and FIR point
      source catalogues. Red squares show the fluxes of PACS 100 and 160\,$\mu$m
      from our photometry measurements. Black circles are fluxes of PACS 160 and
      SPIRE 250, 350, and 500\,$\mu$m from the GCC catalogue of compact sources.
      For the fluxes selected from public point source catalogues, only the source 
      nearest to the GCC source centre was considered. 
      The source number in
      the GCC catalogue is indicated in the top-left corner of SED plot, as well
      as the final classification (SL=starless, PS=protostellar, 1=candidate, 
      2=reliable). In the \Tdust\, profiles, the black points are individual
      pixels, the blue lines show the background \Tdust, the red lines are Gaussian 
      fits, and the grey area show the $1\sigma$ fit uncertainties.
      }
      \label{fig:source_class_example}
   \end{figure*}

   Figure~\ref{fig:source_distrib} shows the distribution of submillimetre compact sources
   from the GCC catalogue. The range of angular sizes (40 - 130 \arcsec) corresponds 
   to approximately 0.15 to 0.50 pc, implying that each source is likely an individual
   dense core, although in some cases the better resolution of short-wavelength 
   maps reveals that the source is substructured. Notable examples are:
   (i) the source 1447 in the GCC catalogue where four sources are resolved 
   at 100\,$\mu$m;
   (ii) the source 1454 where three sources appear in all PACS and WISE maps;
   (iii) the source 1457 where two compact sources are resolved in PACS and 
   SPIRE 250\,$\mu$m maps, one of which is classified as Class 0/I by 
   \citet{rapson_spitzer_2014}.

   The blue and red ellipses in Fig.~\ref{fig:source_distrib} show the starless
   and protostellar sources, respectively. The individual classification is
   given in Table~\ref{tab:stateclass}. Overall, we obtained 78 starless
   cores (21 candidates + 57 reliable), 12 protostellar cores (4 candidates +
   8 reliable), 1 galaxy, and 7 remain unclassified, while in \citet
   {montillaud_galactic_2015} these numbers were 33 (33 + 0), 19 (15 + 4), 0, and 
   46, respectively. Most differences come from unclassified sources for which
   our more careful analysis enabled us to propose a classification. Figure~\ref
   {fig:source_class_example} shows the example of s1458, where the morphology
   of the mid-IR emission reveals that the fluxes measured in WISE bands are not
   due to a point source associated with the core. Interestingly, this source 
   matches a compact source of \NtHp\, emission (Fig.~\ref{fig:source_class_example}, 
   left), confirming that s1458 is a genuine starless dense core. The 
   figure also shows s1446, with a typical protostellar SED and a 4.2 $\pm$ 0.8 K
   increase in \Tdust. Finally s1449 is classified as a candidate protostellar
   source because of its SED, despite a significant decrease in \Tdust\,(-2.1 
   $\pm$ 0.4 K) which is probably due to the envelope of the protostar. We found
   the latter situation to be widespread and therefore excluded the temperature
   profiles from the classification scheme (Sect.~\ref{sec:GCCsources_method}).

   We computed the virial masses of the 33 GCC sources in the IRAM map from both 
   \tCO\, and \CeO\, data. The masses $M^{13}$ and $M^{18}$ of \tCO\, and \CeO\, 
   emitting  gas, respectively, were also computed using the abundances 
   presented in Sect.~\ref{sec:abundances}. Table\,\ref{tab:virial} summarises 
   the characteristics of the 26 sources (totalling 35 velocity components)
   where the peak emission is determined with a SNR$>$3 for at least one of the
   three isotopologues. We computed the virial parameter $\alpha_{\rm vir}^{13} 
   = M_{\rm vir}^{13}/M_{\rm dust}$, where $M_{\rm dust}$ is the mass estimate 
   from the GCC catalogue inferred from dust emission. We similarly computed 
   $\alpha_{\rm vir}^{18}$. These values should be used carefully, since in most
   cases the line widths of the Gaussian fits are such that $\sigma^{12}>\sigma^{13}>\sigma^{18}$
   suggesting that, in our sample, optical depth effects often contribute to the 
   line widths and therefore to the virial mass (over-)estimates. In addition,
   the fact that \CeO\, probes denser layers than \tCO\,can contribute to
   differences in linewidths. Also, when several velocity components
   are present, $M_{\rm dust}$ is over-estimated since it includes the emission
   from all the components.

   In our sample, counting each velocity component individually, 15 sources have
   $\alpha_{\rm vir}^{13}\lesssim 2$. Except for s1461, s1463, s1469, and s1471,
   all those sources correspond to peaks in \NtHp\, emission with integrated 
   intensities 1.2 $< W <$ 6.2 K\,km\,s$^{-1}$. Among the 20 unbound sources, 
   only s1459, s1464, and s1472 correspond to a genuine \NtHp\, emission peak, 
   with $W$=1.0, 1.5, 1.4 K\,km\,s$^{-1}$, respectively. Hence, the overall 
   correlation between \NtHp emission and boundedness is very good as expected
   considering the \NtHp\,(1-0) critical density \citep[$3\times 10^5$ cm$^{-3}$, ][]
   {sanhueza_chemistry_2012}. Seven of the bound sources are classified as 
   protostellar, six as starless, and the two components of s1450 are undetermined. 
   In the 20 unbound sources, one is protostellar, 18 are starless, and one is 
   unclassified, which is broadly consistent with a correlation between the source
   boundedness and its evolutionary stage. The virial parameters $\alpha_{\rm vir}^{18}$
   are generally lower than $\alpha_{\rm vir}^{13}$, but the trends remain similar.

   \begin{sidewaystable*}
   \begin{center}
   \caption{ Fit parameters and virial masses of sources with a peaked emission in \CO, \tCO\, or \CeO.}
   \begin{tabular}{*{19}{c}}
   \toprule
     ID & Comp. & $V_{\rm LSR}$ & $T_{\rm mb}^{12}$ & $T_{\rm mb}^{13}$ & $T_{\rm mb}^{18}$ & $\sigma^{12}$ & $\sigma^{13}$ & $\sigma^{18}$ & $N^{13}$ & $N^{18}$ & $M_{\rm vir}^{13}$ & $M_{\rm vir}^{18}$ & $M^{13}$ & $M^{18}$ & $M_{\rm dust}$ & $\alpha_{vir}^{13}$ & $\alpha_{vir}^{18}$ & Evol.\\
 &  & [km/s]  & [K]  & [K]  & [K]  & [km/s]  & [km/s]  & [km/s]  & [cm$^{-2}$]  & [cm$^{-2}$]  & [$M_\odot$]  & [$M_\odot$]  & [$M_\odot$]  & [$M_\odot$]  & [$M_\odot$]  &  &  &  \\
   \midrule
   1446 & 1 &  8.27 & 16.44 &  7.12 &  0.98 &  1.00 &  0.60 &  0.50 & 1.7e+16 & 1.7e+15 &  19.4 &  13.7 &   7.8 &   6.3 &  21.6 &   0.9 &   0.6 & 4 \\ 
        & 2 & 10.00 &  4.00 &  0.39 &  0.10 &  4.00 &  1.02 &  0.60 & 1.0e+15 & 1.5e+14 &  56.0 &  19.7 &   0.4 &   0.5 &  21.6 &   2.6 &   0.9 & 4 \\ 
   1448 & 1 &  7.86 & 12.27 &  8.45 &  1.22 &  0.94 &  0.55 &  0.48 & 2.0e+16 & 1.8e+15 &  12.3 &   9.4 &   7.2 &   5.4 &  15.3 &   0.8 &   0.6 & 4 \\ 
   1449 & 1 &  7.56 & 14.15 &  8.36 &  1.51 &  0.51 &  0.47 &  0.36 & 1.6e+16 & 1.8e+15 &  11.3 &   6.5 &   7.0 &   6.1 &  23.0 &   0.5 &   0.3 & 4 \\ 
   1450 & 1 &  5.81 & 12.97 &  6.21 &  0.89 &  1.80 &  0.85 &  0.62 & 2.0e+16 & 1.7e+15 &  53.2 &  28.3 &  11.8 &   8.4 &  51.7 &   1.0 &   0.5 & 0 \\ 
        & 2 &  7.44 & 11.02 &  8.11 &  1.34 &  0.95 &  0.48 &  0.41 & 1.6e+16 & 1.6e+15 &  16.7 &  12.1 &   9.6 &   7.9 &  51.7 &   0.3 &   0.2 & 0 \\ 
   1451 & 1 &  5.94 & 16.50 &  8.25 &  1.57 &  0.66 &  0.65 &  0.54 & 2.3e+16 & 3.0e+15 &  25.8 &  18.1 &  11.4 &  12.3 &  16.3 &   1.6 &   1.1 & 3 \\ 
        & 2 &  7.57 & 23.32 & 10.69 &  1.06 &  0.64 &  0.46 &  0.40 & 2.5e+16 & 1.8e+15 &  12.9 &   9.8 &  12.5 &   7.4 &  16.3 &   0.8 &   0.6 & 3 \\ 
   1453 & 1 &  8.53 & 17.33 &  8.15 &  1.81 &  0.70 &  0.43 &  0.32 & 1.5e+16 & 2.1e+15 &  10.3 &   5.5 &   6.8 &   7.7 &  22.3 &   0.5 &   0.2 & 2 \\ 
   1454 & 1 &  5.29 &  9.41 &  6.74 &  2.71 &  1.13 &  0.51 &  0.34 & 1.3e+16 & 2.8e+15 &  17.7 &   8.0 &   7.4 &  12.7 &  32.0 &   0.6 &   0.2 & 4 \\ 
   1459 & 1 &  7.71 & 18.38 & 10.58 &  1.91 &  1.02 &  0.48 &  0.31 & 2.4e+16 & 2.2e+15 &  11.2 &   4.8 &   9.9 &   7.6 &   4.4 &   2.5 &   1.1 & 1 \\ 
   1460 & 1 &  5.81 &  5.31 &  1.44 &  0.06 &  0.46 &  0.56 &  1.42 & 2.2e+15 & 2.0e+14 &  15.3 &  98.9 &   0.9 &   0.7 &   8.5 &   1.8 &  11.6 & 2 \\ 
        & 2 &  7.99 & 15.00 &  8.56 &  3.30 &  1.25 &  0.49 &  0.22 & 1.8e+16 & 2.6e+15 &  11.8 &   2.4 &   7.3 &   8.7 &   8.5 &   1.4 &   0.3 & 2 \\ 
   1461 & 1 &  4.98 &  5.77 &  2.64 &  0.67 &  0.22 &  0.22 &  0.33 & 1.7e+15 & 5.6e+14 &   4.1 &   9.4 &   1.3 &   3.4 &  13.3 &   0.3 &   0.7 & 2 \\ 
        & 2 &  5.65 & 12.35 &  5.66 &  2.08 &  0.59 &  0.59 &  0.36 & 1.2e+16 & 2.4e+15 &  30.5 &  11.2 &   8.9 &  14.2 &  13.3 &   2.3 &   0.8 & 2 \\ 
   1463 & 1 &  5.47 & 12.93 &  6.07 &  1.76 &  0.45 &  0.45 &  0.32 & 1.0e+16 & 1.8e+15 &   8.7 &   4.6 &   3.9 &   5.7 &   4.7 &   1.9 &   1.0 & 2 \\ 
   1464 & 1 &  5.92 &  9.29 &  5.70 &  0.92 &  0.87 &  0.82 &  0.54 & 1.7e+16 & 1.4e+15 &  34.3 &  14.8 &   7.2 &   4.8 &   2.9 &  11.8 &   5.1 & 1 \\ 
        & 2 &  8.45 & 17.48 &  6.64 &  0.81 &  0.72 &  0.36 &  0.24 & 9.7e+15 & 6.9e+14 &   6.6 &   3.0 &   4.2 &   2.4 &   2.9 &   2.3 &   1.0 & 1 \\ 
   1469 & 1 &  5.43 &  7.04 &  5.27 &  2.47 &  2.03 &  0.49 &  0.26 & 9.6e+15 & 1.9e+15 &  16.9 &   5.0 &   5.6 &   8.9 &  13.7 &   1.2 &   0.4 & 2 \\ 
   1470 & 1 &  8.28 & 18.92 &  8.88 &  1.08 &  0.95 &  0.43 &  0.31 & 1.7e+16 & 1.3e+15 &   9.9 &   5.1 &   7.6 &   4.5 &   2.8 &   3.5 &   1.8 & 1 \\ 
   1471 & 1 &  7.31 &  6.38 &  5.67 &  0.96 &  1.70 &  0.31 &  0.14 & 7.6e+15 & 3.6e+14 &   4.8 &   1.0 &   3.2 &   1.2 &   2.9 &   1.7 &   0.3 & 4 \\ 
   1472 & 1 &  6.82 & 16.57 &  8.64 &  1.45 &  1.29 &  0.70 &  0.49 & 2.6e+16 & 2.5e+15 &  26.1 &  13.0 &  11.5 &   9.1 &   8.5 &   3.1 &   1.5 & 1 \\ 
   1477 & 1 &  5.21 &  4.23 &  2.14 &  0.96 &  0.45 &  0.45 &  0.26 & 2.9e+15 & 6.6e+14 &  17.7 &   5.9 &   2.2 &   4.0 &   2.2 &   8.0 &   2.7 & 2 \\ 
   1480 & 1 &  8.14 & 15.21 &  8.63 &  2.11 &  1.38 &  0.59 &  0.30 & 2.1e+16 & 2.2e+15 &  12.8 &   3.3 &   7.6 &   6.2 &   2.8 &   4.6 &   1.2 & 2 \\ 
   1484 & 1 &  5.87 & 10.85 &  4.39 &  0.79 &  0.89 &  0.57 &  0.26 & 8.3e+15 & 6.0e+14 &  14.4 &   3.1 &   3.3 &   1.9 &   2.2 &   6.5 &   1.4 & 0 \\ 
   1486 & 1 &  7.85 & 16.64 &  7.87 &  0.69 &  0.71 &  0.41 &  0.35 & 1.4e+16 & 8.4e+14 &  10.5 &   7.6 &   6.9 &   3.5 &   1.2 &   8.8 &   6.3 & 2 \\ 
   1491 & 1 &  9.13 & 16.24 &  3.10 &  0.10 &  0.62 &  0.39 &  0.25 & 4.4e+15 & 8.0e+13 &  11.0 &   4.4 &   2.6 &   0.4 &   2.4 &   4.6 &   1.8 & 2 \\ 
   1495 & 1 &  6.49 &  9.75 &  4.99 &  0.83 &  1.03 &  0.75 &  0.48 & 1.3e+16 & 1.1e+15 &  36.0 &  15.0 &   6.7 &   4.8 &   1.5 &  24.0 &  10.0 & 2 \\ 
        & 2 &  9.12 & 15.39 &  6.38 &  0.53 &  0.79 &  0.43 &  0.34 & 1.1e+16 & 6.0e+14 &  11.8 &   7.6 &   5.6 &   2.6 &   1.5 &   7.9 &   5.1 & 2 \\ 
   1514 & 1 &  6.55 &  7.11 &  5.48 &  0.90 &  2.08 &  0.54 &  0.27 & 1.1e+16 & 6.5e+14 &  14.3 &   3.7 &   4.7 &   2.2 &   1.2 &  11.9 &   3.1 & 2 \\ 
        & 2 &  8.82 & 11.62 &  7.25 &  0.83 &  0.72 &  0.46 &  0.34 & 1.3e+16 & 8.5e+14 &  10.6 &   5.9 &   5.5 &   2.9 &   1.2 &   8.8 &   4.9 & 2 \\ 
   1518 & 1 &  5.40 & 11.10 &  7.42 &  1.58 &  1.13 &  0.56 &  0.39 & 1.6e+16 & 1.9e+15 &  14.9 &   7.3 &   6.7 &   6.2 &   1.6 &   9.3 &   4.6 & 2 \\ 
   1521 & 1 &  5.74 &  9.58 &  3.39 &  0.36 &  1.27 &  0.57 &  0.28 & 6.0e+15 & 2.8e+14 &  23.1 &   5.5 &   3.5 &   1.3 &   1.4 &  16.5 &   3.9 & 2 \\ 
        & 2 &  8.76 & 16.70 &  7.39 &  0.56 &  0.50 &  0.35 &  0.32 & 1.1e+16 & 6.3e+14 &   9.0 &   7.5 &   6.3 &   3.0 &   1.4 &   6.4 &   5.4 & 2 \\ 
   1528 & 1 &  5.49 & 14.71 &  3.92 &  0.28 &  0.63 &  0.41 &  0.26 & 5.7e+15 & 2.3e+14 &   9.5 &   3.8 &   2.6 &   0.9 &   0.6 &  15.8 &   6.3 & 2 \\ 
        & 2$^\star$ &  5.41 &  0.12 &  1.73 &  0.18 &  0.37 &  0.15 &  0.03 & 6.6e+15$^\dagger$ & 2.3e+14$^\dagger$ &   - &   - &   3.3$^\dagger$ &   1.0$^\dagger$ &   0.6 &  - & - & 2 \\ 
   \bottomrule
   \end{tabular}
   \tablefoot{The columns are:
   (1) Index of the source in the GCC catalogue; 
   (2) Index of the velocity component; 
   (3) Centroid velocity of the velocity component;
   (4-6) Peak \Tmb\, of the Gaussian fit to the current velocity component of the \CO, \tCO, and \CeO\, lines;
   (7-9) Standard deviation of the Gaussian fit to the current velocity component of the \CO, \tCO, and \CeO\, lines;
   (10-11) Column density of \tCO\, and \CeO;
   (12-13) Virial mass from \tCO\,and \CeO\, line widths;
   (14-15) Core mass from \tCO\, and \CeO;
   (16) Estimates of the core mass from the GCC catalogue;
   (17-18) Virial parameters from \tCO\, and \CeO\,(1-0) lines;
   (19) New estimate of the evolutionary stage of the source. 
      The code is: 
      0=undetermined,
      1=candidate starless core,
      2=reliable starless core,
      3=candidate protostellar core,
      4=reliable protostellar core.
   $\star$ This component corresponds to the non-Gaussian shape of the first component peak, hence the very small value of $T_{\rm mb}^{12}$. The quantities marked with $\dagger$ were derived from a full integration of the line profile, therefore including both components.
   }
   \label{tab:virial}
   \end{center}
   \end{sidewaystable*}

\section{Discussion: A peak in star formation activity}\label{sec:discussion} 

   The star formation activity can be characterised by several tracers. \citet
   {rapson_spitzer_2014} classified all the {\it Spitzer} sources around NGC 2264
   up to the region of the northern filaments of G202.3+2.5. Their Fig.~1 shows
   that the surface density of protostars decreases when moving from NGC 2264 to
   G202.3+2.5, and that younger objects (mostly Class 0/I) are found in G202.3+2.5
   than in NGC 2264 (mostly Class II). This could support the idea of a gradient 
   and/or a regular time sequence in star formation activity along the south-north 
   direction. Alternatively, the younger sources observed in G202.3+2.5 could be
   the product of a new episode of star formation in this region.

   The distribution of masses from dust observations reported in Table
   \ref{tab:virial} reveals that six of the nine sources with $M_{\rm dust}>10 M_\odot$,
   including the $52 M_\odot$ core s1450 where no firm signs of protostellar
   activity were found, form a chain along the \NtHp-bright ridge
   (Fig.~\ref{fig:source_class_example}) at the junction between the three main 
   filaments of the ramified structure of the cloud (Fig.~\ref{fig:source_distrib}, 
   hereafter 'the junction region'). Interestingly, the ridge also hosts
   the massive protostellar core s1446 (=NGC 2264 H), where \citet
   {wolf-chase_star_2003}, using \CO\,(2-1) observations, have detected an 
   outflow associated with the nearby Herbig-Haro objects HH 576 and HH 577.
   We further investigate this outflow in the next paper of this series. Adding
   only the massive source masses, the ridge is at least $150 M_\odot$. The 
   second most massive source in Table~\ref{tab:virial} is s1454 ($32 M_\odot$) 
   where a secondary \NtHp\,peak is observed (hereafter 'the north clump').

\begin{figure}
      \includegraphics[width=0.5\textwidth]{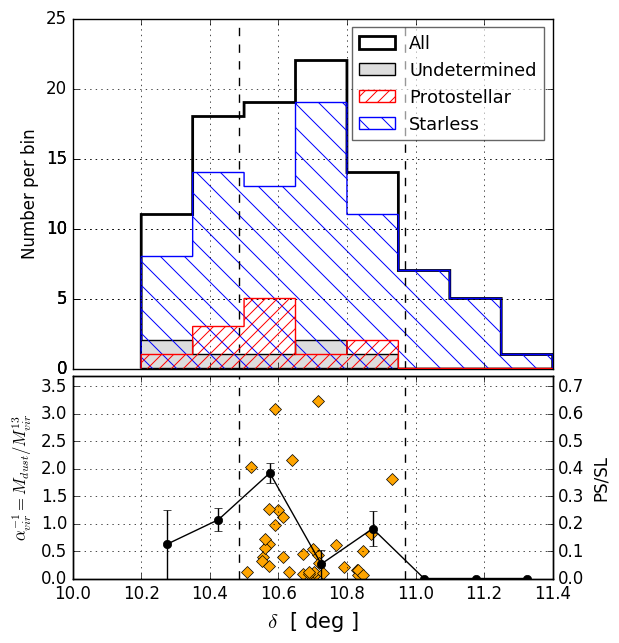}
      \caption{Distributions of submillimetre sources as a function of declination.
      The blue histogram is for starless cores, the red histogram for protostellar
      cores, the grey one for sources with undetermined stage of evolution, and 
      the black one for all sources. In the lower panel, the black circles show the ratio 
      between the numbers of protostellar and starless cores in each bin, and 
      the orange diamonds show the ratio $M_{\rm dust}/M_{\rm vir}^{13}$ for
      individual sources. The vertical dashed lines show the limits of the area
      mapped with the IRAM 30-m telescope.
      }
      \label{fig:SFgradient}
   \end{figure}

   We use our source classification to estimate the star formation activity as a
   function of declination, a good measure of the distance to the open cluster. 
   Figure~\ref{fig:SFgradient} shows the variations in the number of sources 
   with declination. The distribution peaks near $\delta = 10.7^\circ$, 
   slightly north of the junction region. The distribution of protostellar sources
   peaks near $\delta = 10.6^\circ$, at the junction region, where most of the 
   \NtHp-bright sources (Fig.~\ref{fig:source_class_example}) are located. The 
   protostellar-to-starless source ratio also peaks in the junction region, and 
   presents a secondary peak at $\delta = 10.9^\circ$, near the north clump. 
   
   These trends are strengthened by the distribution of the virial parameter 
   $\alpha_{\rm vir}^{13}$, where ten out of the 15 bound sources ($\alpha_{\rm vir}^{-1} 
   > 0.5$ in Fig.~\ref{fig:SFgradient}) are located in the junction region 
   ($10.5^\circ < \delta < 10.7^\circ$), three are near the north clump ($10.8^\circ 
   < \delta < 11.0^\circ$), and the unbound sources are evenly distributed between 
   $\delta = 10.5^\circ$ and $10.9^\circ$.

   Hence, our results strongly suggest a local increase in star formation activity
   near s1446, consistent with the scenario of a new episode of star formation
   in this region at the northern outskirts of the Mon OB1 complex.

\section{Conclusion and perspectives}\label{sec:conclusion} 

We have examined the core-scale dust and gas observables in the cloud
G202.3+2.5. The evolutionary classification of compact sources was 
improved compared to \citet{montillaud_galactic_2015} to analyse the star
formation activity. We have shown that the 
junction region in G202.3+2.5 is: (i) a massive and dense ridge hosting several 
massive cores, including the active source 1446, responsible for an outflow, and
the possibly starless 52 $M_\odot$ core s1450; (ii) a local 
peak in the star formation activity of the Mon OB 1 complex. It is striking that
this region lies at the junction of three branches of the ramified structure of 
G202.3+2.5, suggesting that these branches follow a convergent dynamics that has
enhanced the star formation activity in the junction region. Determining the 
origin of this enhancement now requires to extend the analysis to larger scales, 
to examine the relationship between cores, filaments and their environment. This 
multi-scale analysis is the focus of the next paper in this series.

\begin{acknowledgements}

   This work is based on observations carried out under project number 113-16 
   with the IRAM 30-m telescope. IRAM is supported by INSU/CNRS (France), MPG 
   (Germany) and IGN (Spain). JMo warmly thanks the staff of the 30m for its
   kind and efficient help, and in particular Pablo Garcia for stimulating
   discussions.\\

   The project leading to this publication has received funding from the 'Soutien 
   à la recherche de l'observatoire' by the OSU THETA.\\

   This work was supported by the Programme National “Physique et Chimie du Milieu 
   Interstellaire” (PCMI) of CNRS/INSU with INC/INP co-funded by CEA and CNES.\\
   
   JMo, RB, DC, and VLT thank the French ministry of foreign affairs (French 
   embassy in Budapest) and the Hungarian national office for research and 
   innovation (NKFIH) for financial support (Balaton program 40470VL/ 
   2017-2.2.5-TÉT-FR-2017-00027).\\

   The project leading to this publication has received funding from the European 
   Union's Horizon 2020 research and innovation programme under grant agreement 
   No 730562 [RadioNet]\\

   This research used data from the Second Palomar Observatory Sky Survey (POSS-II),
   which was made by the California Institute of Technology with funds from the 
   National Science Foundation, the National Geographic Society, the Sloan 
   Foundation, the Samuel Oschin Foundation, and the Eastman Kodak Corporation.\\

   MJ and ERM acknowledge the support of the Academy of Finland Grant No. 285769.\\

   This research has made use of the SVO Filter Profile Service (http://svo2.cab.inta-csic.es/theory/fps/) 
   supported from the Spanish MINECO through grant AyA2014-55216.\\

   J.H.He is supported by the NSF of China under Grant Nos. 11873086 and 
   U1631237, partly by Yunnan province (2017HC018), and also partly by the 
   Chinese Academy of Sciences (CAS) through a grant to the CAS South America 
   Center for Astronomy (CASSACA) in Santiago, Chile.

   CWL was supported by the Basic Science Research Program through the National 
   Research Foundation of Korea (NRF) funded by the Ministry of Education, 
   Science and Technology (NRF-2019R1A2C1010851).

   SZ acknowledge the support of NAOJ ALMA Scientific Research Grant Number 2016-03B.

   KW acknowledges support by
   the National Key Research and Development Program of China (2017YFA0402702),
   the National Science Foundation of China (11973013, 11721303),
   and the starting grant at the Kavli Institute for Astronomy and Astrophysics, Peking University (7101502016).

   This work has made use of data from the European Space Agency (ESA) mission
   {\it Gaia} (\url{https://www.cosmos.esa.int/gaia}), processed by the {\it Gaia}
   Data Processing and Analysis Consortium (DPAC,
   \url{https://www.cosmos.esa.int/web/gaia/dpac/consortium}). Funding for the DPAC
   has been provided by national institutions, in particular the institutions
   participating in the {\it Gaia} Multilateral Agreement.
\end{acknowledgements}

\longtab{
   \begin{landscape}
   \begin{longtable}{ c c c c c c c c c c l }
   \caption{\label{tab:stateclass} Evolutionary state classification of the submillimetre compact sources.}\\
   \toprule
ID & Type & $N($H$_2$) & $T_d$ & $\Delta T_d$  & w1-w2 & w3-w4 & Type & Gaia match & $\pi$ & Comment\\
   &      & (cm$^{-2}$) & (K)  & (K)           &       &       &      &            & (mas) & \\
   \midrule
   \endfirsthead
   \caption{continued.}\\
   \toprule
ID & Type & $N($H$_2$) & $T_d$ & $\Delta T_d$  & w1-w2 & w3-w4 & Type & Gaia match & $\pi$ & Comment\\
   &      & (cm$^{-2}$) & (K)  & (K)           &       &       &      &            & (mas) & \\
   \midrule
   \endhead
   \bottomrule
   \hline
   \endfoot
1446 & 4 & 4.4(22) & 17.1 & 4.2 $\pm$ 0.8 & cp & cp & 0/I & - & - & Outflows, HH objects\\
1447 & 4 & 8.7(22) & 13.1 & -1.3 $\pm$ 0.9 & cp & cp & 0/I & - & - & Several YSOs, outflows, HH objects\\
1448 & 4 & 3.8(22) & 14.1 & -0.2 $\pm$ 0.9 & cp & cp & 0/I & - & - & Source detected at 3-500 $\mu$m\\
1449 & 4 & 5.0(22) & 10.2 & -2.1 $\pm$ 0.3 & cp & cp & 0/I & - & - & Source detected at 3-500 $\mu$m\\
1450 & 0 & 7.9(22) & 9.9 & -2.0 $\pm$ 0.4 & cp & bg & III/F & 3351030967214777984 & 1.45 $\pm$ 0.05 & 100 $\mu$m emission : Possibly one Class 0 YSO\\
1451 & 3 & 2.9(22) & 12.7 & -0.8 $\pm$ 0.4 & cp & cp & III/F & - & - & Source detected at 3-500 $\mu$m $\neq$ Rapson\\
1452 & 4 & 2.7(22) & 13.0 & -0.9 $\pm$ 0.4 & cp & cp & II & 3350981656693549696 & 1.76 $\pm$ 0.36 & Source detected at 3-500 $\mu$m\\
1453 & 2 & 4.5(22) & 9.4 & -2.6 $\pm$ 0.4 & bg & bg & AGN & - & - & Large spatial shift\\
1454 & 4 & 5.2(22) & 9.8 & -2.2 $\pm$ 0.4 & cp & cp & 0/I & - & - & Source detected at 3-500 $\mu$m – Several YSOs\\
1455 & 0 & 2.9(22) & 10.1 & -1.9 $\pm$ 0.3 & cp & bg & III/F & 3326958152260041600 & 1.7 $\pm$ 0.2 & NIR source is CIII/F, but compact source at 100 $\mu$m\\
1456 & 1 & 2.4(22) & 12.2 & -1.9 $\pm$ 0.2 & cp & bg & III/F & - & - & \\
1457 & 4 & 2.0(22) & 11.7 & -0.7 $\pm$ 0.2 & cp & cp & 0/I & - & - & Detected at 3-500 $\mu$m – Substructured, at least one YSO\\
1458 & 2 & 1.4(22) & 11.4 & -1.2 $\pm$ 0.3 & bg & - & - & - & - & \\
1459 & 1 & 9.6(21) & 12.2 & -0.8 $\pm$ 0.3 & bg & bg & III/F & 3350983683919806848 & 3.29 $\pm$ 0.87 & 2MASS=Gaia $\neq$ Rapson\\
1460 & 2 & 1.9(22) & 11.1 & -1.3 $\pm$ 0.3 & cp & - & III/F & 3351030417458975616  & 1.98 $\pm$ 0.03 & \\
1461 & 2 & 1.6(22) & 11.7 & -1.2 $\pm$ 0.4 & - & - & - & - & - & \\
1462 & 2 & 1.1(22) & 11.0 & -1.0 $\pm$ 0.4 & - & bg & - & - & - & \\
1463 & 2 & 1.1(22) & 11.4 & -0.9 $\pm$ 0.2 & - & - & - & - & - & \\
1464 & 1 & 6.2(21) & 11.9 & -0.6 $\pm$ 0.1 & cp & bg & III/F & 3351033883496495744 & 1.45 $\pm$ 0.14 & 2MASS=Gaia $\neq$ Rapson\\
1465 & 2 & 1.1(22) & 11.7 & -1.5 $\pm$ 0.3 & - & bg & - & - & - & \\
1466 & 1 & 6.3(21) & 13.9 & -0.6 $\pm$ 0.2 & cp & bg & - & 3350978602973521280 & 3.4 $\pm$ 0.7 & \\
1467 & 0 & 1.5(22) & 11.9 & -1.0 $\pm$ 0.3 & cp & bg & III/F & - & - & Several unclassified NIR/MIR point sources\\
1468 & 3 & 7.5(21) & 13.7 & -0.3 $\pm$ 0.3 & cp & cp & III/F & 3351046394737185408 & 1.29 $\pm$ 0.05 & Source detected at 3-500 $\mu$m – real CIII ?\\
1469 & 2 & 2.1(22) & 9.4 & -1.8 $\pm$ 0.2 & cp & bg & III/F & 3351056973241029248 & 1.9 $\pm$ 0.04 & \\
1470 & 1 & 5.8(21) & 13.5 & -0.7 $\pm$ 0.2 & bg & bg & - & - & - & \\
1471 & 4 & 6.3(21) & 12.3 & -0.8 $\pm$ 0.2 & cp & cp & 0/I & - & - & Several foreground/background sources, one real YSO\\
1472 & 1 & 1.7(22) & 8.8 & -0.6 $\pm$ 0.2 & cp+bg & bg & III/F & - & - & \\
1473 & 2 & 6.3(21) & 12.0 & -1.2 $\pm$ 0.2 & bg & - & - & - & - & \\
1474 & 2 & 1.3(22) & 12.0 & -0.6 $\pm$ 0.4 & - & bg & - & - & - & \\
1475 & 2 & 4.4(21) & 12.7 & -0.6 $\pm$ 0.2 & cp & - & III/F & 3326959011253582848 & 1.2 $\pm$ 0.7 & Large spatial shift\\
1476 & 2 & 1.4(22) & 9.9 & -0.3 $\pm$ 0.3 & cp & - & III/F & - & - & Large spatial shift\\
1477 & 2 & 2.7(21) & 14.0 & -0.7 $\pm$ 0.8 & cp & bg & III/F & - & - & \\
1478 & 2 & 6.3(21) & 12.2 & -0.7 $\pm$ 0.4 & cp & - & III/F & 3351040858523618816 & 1.19 $\pm$ 0.24 & Gaia = Rapson, large spatial fit\\
 &  &  &  &    &  &  &  & 3351040858524306944 & 1.26 $\pm$ 0.37 & Gaia = 2MASS, large spatial fit\\
1479 & 2 & 6.3(21) & 12.7 & -1.0 $\pm$ 0.2 & cp & - & - & - & - & Large spatial shift\\
1480 & 2 & 7.3(21) & 11.9 & -0.9 $\pm$ 0.3 & cp & - & III/F & 3351030795416094848 & 1.76 $\pm$ 0.11 & \\
1481 & 3 & 4.7(21) & 13.7 & -0.7 $\pm$ 0.2 & cp & cp & II & 3326961794392311424 & 1.62 $\pm$ 0.39 & Missing compact emission at 100 $\mu$m\\
1482 & 2 & 7.5(21) & 11.4 & -0.9 $\pm$ 0.4 & cp & - & III/F & 3350981862852157056 & 0.36 $\pm$ 0.29 & Large spatial shift\\
1483 & 2 & 5.7(21) & 10.4 & -1.4 $\pm$ 0.2 & cp & - & III/F & 3351032101084459776 & 0.6 $\pm$ 1 & Large spatial shift\\
1484 & 0 & 5.1(21) & 12.6 & -0.4 $\pm$ 0.1 & cp & - & III/F &  335105645784390912 & -1.6 $\pm$ 0.8 & Type=CIII/F, but compact emission at 100$\mu$m\\
1485 & 2 & 1.3(22) & 10.1 & -1.9 $\pm$ 0.3 & cp & bg & III/F & 3351089000812448640 & 1.69 $\pm$ 0.09 & 2MASS = Rapson $\neq$ Gaia\\
1486 & 2 & 2.2(21) & 18.2 & 0.3 $\pm$ 0.4 & cp & cp & III/F & 3350983069739510528 & 0.2 $\pm$ 0.8 & \\
1487 & 1 & 2.6(21) & 13.9 & -1.1 $\pm$ 0.1 & cp+bg & bg & - & - & - & \\
1488 & 2 & 3.1(21) & 12.7 & -0.3 $\pm$ 0.1 & - & bg & - & - & - & \\
1489 & 2 & 2.9(21) & 15.3 & 0.6 $\pm$ 0.2 & cp & bg & III/F & - & - & Probable warm starless core\\
1490 & 0 & 3.0(21) & 14.4 & -0.8 $\pm$ 0.2 & cp & bg & - & 3327782889059332480 & 1.37 $\pm$ 0.2 & Good spatial and distance match, but no MIR detection\\
1491 & 2 & 3.8(21) & 15.1 & -0.1 $\pm$ 0.1 & cp & bg & III/F & 3351038453342829952 & - & Large spatial shift\\
1492 & 1 & 5.9(21) & 10.8 & -0.8 $\pm$ 0.2 & cp & bg & III/F & 3351043710382744832 & - & Substructured\\
1493 & 2 & 1.1(22) & 10.1 & -1.0 $\pm$ 0.2 & bg & bg & - & - & - & \\
1494 & 0 & 1.5(21) & 14.8 & -0.5 $\pm$ 0.2 & cp & cp & - & 3327031995634205952 & 0.27 $\pm$ 0.05 & 2MASS=Gaia (background), but 100 $\mu$m emission\\
1495 & 2 & 2.6(21) & 13.2 & 0.1 $\pm$ 0.2 & cp & bg & III/F & - & - & Core-shaped emission at 100 $\mu$m\\
1496 & 1 & 3.4(21) & 11.5 & -0.5 $\pm$ 0.2 & cp & bg & PAH & - & - & \\
1497 & 3 & 8.0(20) & 15.9 & 0.1 $\pm$ 0.4 & cp & cp & 0/I & - & - & YSO candidate not detected at 100 $\mu$m\\
1498 & 2 & 3.0(21) & 15.0 & -0.3 $\pm$ 0.2 & cp & bg & III/F & 3351057522995678208  & 0.35 $\pm$ 0.22 & \\
1499 & 2 & 2.4(21) & 14.7 & 0.5 $\pm$ 0.1 & cp & core & III/F & 3351061955401965952 & 1.1 $\pm$ 0.1 & Warm starless – Core-shaped emission at 12-160 $\mu$m\\
1500 & 1 & 3.4(21) & 11.3 & -0.9 $\pm$ 0.2 & cp & - & III/F & - & - & \\
1501 & 1 & 3.2(21) & 15.0 & 0.1 $\pm$ 0.2 & cp & - & III/F & - & - & \\
1502 & 2 & 3.7(21) & 11.3 & -0.5 $\pm$ 0.3 & cp & - & III/F & 3351040514926184960 & 0.21 $\pm$ 0.15 & \\
1503 & 2 & 3.5(21) & 11.9 & -0.2 $\pm$ 0.1 & bg & bg & - & - & - & \\
1504 & 2 & 2.0(21) & 14.1 & -0.2 $\pm$ 0.1 & cp & bg & - & 3350978564318784896 & 4.6 $\pm$ 0.6 & \\
1505 & 0 & 4.9(21) & 13.0 & -0.3 $\pm$ 0.2 & cp & bg & III/F & 3326946233727400704 & 2 $\pm$ 0.2 & 2MASS = Rapson = Gaia, foreground\\
 &  &  &  &    &  &  &  & 3326946233727399680 & 1.27 $\pm$ 0.26 & Second Gaia source close to core peak\\
1506 & 2 & 2.4(21) & 13.7 & -0.5 $\pm$ 0.1 & cp & bg & III/F & 3351051475681700864 & -0.1 $\pm$ 0.3 & \\
1507 & 1 & 4.2(21) & 15.1 & -0.4 $\pm$ 0.2 & cp & bg & III/F & 3351045604463218304 & 1.9 $\pm$ 0.2 & \\
1508 & 2 & 4.8(21) & 12.4 & -0.2 $\pm$ 0.1 & bg & bg & - & - & - & \\
1509 & 9 & 8.4(20) & 15.8 & -0.1 $\pm$ 0.2 & cp & cp & - & - & - &  Galaxy: SED peak at 50 $\mu$m, very isolated from filament\\
1510 & 2 & 4.1(21) & 11.4 & -0.4 $\pm$ 0.1 & cp & - & - & 3326944648882916352  & 1.7 $\pm$ 0.6 & Large spatial shift\\
1511 & 2 & 1.7(21) & 16.8 & -0.6 $\pm$ 0.2 & cp & cp & - & 3351857387641672064 & 3.7 $\pm$ 1.3 & The foreground point source is still visible at 22 $\mu$m\\
1512 & 2 & 4.2(21) & 13.0 & -0.3 $\pm$ 0.3 & - & - & - & - & - & \\
1513 & 2 & 1.2(21) & 17.8 & 0.2 $\pm$ 0.1 & cp & core & III/F & 3350984268035096704 & 0.3 $\pm$ 0.3 & Warm starless – Core-shaped emission at 12-160 $\mu$m\\
 &  &  &  &    & cp &  &  & 3350984925165355904 & 1.33 $\pm$ 0.03 & Large spatial shift\\
1514 & 2 & 2.6(21) & 13.4 & -0.3 $\pm$ 0.1 & cp & bg & III/F & 3351034089655997824 & 0.06 $\pm$ 0.55 & 2MASS = Rapson $\neq$ Gaia\\
1515 & 2 & 3.4(21) & 11.2 & -0.3 $\pm$ 0.1 & - & - & - & - & - & \\
1516 & 1 & 2.7(21) & 12.4 & -0.1 $\pm$ 0.3 & cp & bg & - & - & - & Bright ridge emission at 12-100 $\mu$m\\
1517 & 2 & 2.7(21) & 11.5 & -0.4 $\pm$ 0.2 & cp & cp+bg & III/F & 3351035360964589440 & -0.7 $\pm$ 0.9 & \\
1518 & 2 & 3.5(21) & 13.4 & -0.1 $\pm$ 0.3 & cp & - & III/F & 3351039965171558912 & - & Large spatial shift\\
1519 & 2 & 1.4(21) & 16.7 & 0.3 $\pm$ 0.1 & cp & edge & III/F & 3351057458573023104 & -0.8 $\pm$ 0.5 & Warm starless – Core-shaped emission at 12-160 $\mu$m\\
1520 & 1 & 4.0(21) & 12.0 & -0.9 $\pm$ 0.1 & cp & - & III/F & - & - & \\
1521 & 2 & 2.2(21) & 13.7 & -0.1 $\pm$ 0.1 & cp & cp & III/F & 3351056492203698304 & 3.15 $\pm$ 0.05 & \\
1522 & 1 & 3.4(21) & 11.4 & -0.3 $\pm$ 0.1 & cp & bg & III/F & 3351084632831098240 & 0.16 $\pm$ 0.49 & \\
1523 & 2 & 1.2(21) & 13.2 & 0.2 $\pm$ 0.1 & cp & bg & III/F & 3350982137731278720 & 0.51 $\pm$ 0.27 & \\
1524 & 2 & 3.2(21) & 11.6 & -0.2 $\pm$ 0.0 & cp & bg & III/F & - & - & Large spatial shift\\
1525 & 2 & 3.0(21) & 11.9 & 0.1 $\pm$ 0.2 & mcp & bg & III/F & - & - & \\
1526 & 2 & 1.9(21) & 15.2 & 0.3 $\pm$ 0.1 & mcp & edge & PAH & - & - & Warm starless – Core-edge shaped emission at 12-100 $\mu$m\\
1527 & 2 & 2.8(21) & 12.7 & -0.7 $\pm$ 0.1 & cp & cp & III/F & 3351063574606062976 & 0.84 $\pm$ 0.11 & \\
1528 & 2 & 1.1(21) & 14.3 & 0.2 $\pm$ 0.2 & cp & bg & - & 3351056664002342784 & 0.48 $\pm$ 0.66 & Warm starless – Core shaped emission at 100 $\mu$m\\
1529 & 2 & 1.2(21) & 13.4 & -0.2 $\pm$ 0.1 & cp & bg & III/F & 3351037942239970304 & 0.6 $\pm$ 0.5 & 2MASS $\neq$ Rapson = Gaia ; large shifts\\
1530 & 1 & 3.3(21) & 13.7 & -0.5 $\pm$ 0.2 & mcp & bg & AGN & - & - & \\
1531 & 1 & 1.2(21) & 14.9 & 0.5 $\pm$ 0.1 & cp+bg & edge & - & - & - & Warm starless – Core shaped emission at 12-22 $\mu$m\\
1532 & 1 & 1.8(21) & 14.4 & -0.4 $\pm$ 0.1 & cp & - & - & - & - & \\
1533 & 2 & 2.3(21) & 13.2 & -0.6 $\pm$ 0.1 & mcp & bg & III/F & - & - & 2MASS $\neq$ Rapson ; large shift for Rapson\\
1534 & 2 & 1.2(21) & 16.2 & 0.2 $\pm$ 0.1 & cp+bg & core & - & 3327028349205172736 & 0.3 $\pm$ 0.5 & Warm starless – Core shaped emission at 12-100 $\mu$m\\
1535 & 2 & 2.1(21) & 13.5 & -0.2 $\pm$ 0.1 & cp & cp & III/F & 3326956885246723328 & 0.7 $\pm$ 0.3 & \\
1536 & 2 & 1.3(21) & 14.5 & -0.3 $\pm$ 0.2 & mcp & bg & AGN & 3351057973968718592 & 2.1 $\pm$ 0.6 & 2MASS = Gaia $\neq$ Rapson\\
1537 & 2 & 1.3(21) & 13.3 & -0.2 $\pm$ 0.1 & cp & cp & III/F & 3351090645784954624 & 0.21 $\pm$ 0.05 & Large spatial shift\\
1538 & 1 & 6.0(21) & 9.6 & -0.1 $\pm$ 0.1 & cp & edge & AGN & 3350964820423096704 & 0.3 $\pm$ 1.1 & 2MASS = Gaia $\neq$ Rapson\\
1539 & 2 & 9.1(20) & 13.5 & -0.1 $\pm$ 0.1 & cp & core & III/F & 3351036941512554752 & 0.6 $\pm$ 1.2 & Core shaped emission at 12-100 $\mu$m\\
1540 & 2 & 5.2(20) & 18.8 & -0.1 $\pm$ 0.1 & cp & core & III/F & 3327010791379052800 & 0.5 $\pm$ 0.3 & Large spatial shift\\
1541 & 1 & 9.2(20) & 13.9 & 0.2 $\pm$ 0.1 & cp & bg & - & - & - & Warm starless – Core shaped emission at 100 $\mu$m\\
1542 & 1 & 1.2(21) & 15.1 & 0.1 $\pm$ 0.1 & mcp & env. & - & 3350964000083713664 & 0.5 $\pm$ 1 & Envelope emission at 12 $\mu$m\\
1543 & 2 & 1.3(21) & 14.6 & -0.1 $\pm$ 0.1 & mcp & bg & III/F & - & - & 2MASS $\neq$ Rapson ; large shifts\\
\end{longtable}
\tablefoot{
The columns are: 
   (1) Index of the source in the GCC catalogue; 
   (2) New estimate of the evolutionary stage of the source. 
      The code is: 
      0=undetermined,
      1=candidate starless core,
      2=reliable starless core,
      3=candidate protostellar core,
      4=reliable protostellar core.
   (3-4) Column density and dust temperature from the GCC catalogue;
   (5) Variation (centre-background) in dust temperature and its uncertainty;
   (6) Morphology type in the WISE 3.4 and 4.6 $\mu$m bands; 
      cp means compact, 
      mcp means that multiple compact sources are detected,
      bg means background,
   (7) Morphology type in the WISE 12 and 22 $\mu$m bands; 
      core means that the emission has the same morphology as the core at 250 $\mu$m; 
      edge means that the emission corresponds to a bright edge of the core as seen at 250 $\mu$m; 
      env. means envelope of the core seen at 250 $\mu$m; 
   (8) Type of the mid-IR source(s) as classified by \citet{rapson_spitzer_2014};
   (9) Identifier of Gaia source(s) that match the near-/mid-IR source(s);
   (10) Parallax of the Gaia sources.
}
\end{landscape}
}

\bibliographystyle{aa}
\bibliography{Article_G202}

\begin{thebibliography}{34}
\expandafter\ifx\csname natexlab\endcsname\relax\def\natexlab#1{#1}\fi

\bibitem[{Cutri {et~al.}(2011)Cutri, Wright, Conrow, Bauer, Benford,
  Brandenburg, Dailey, Eisenhardt, Evans, Fajardo-Acosta, Fowler, Gelino,
  Grillmair, Harbut, Hoffman, Jarrett, Kirkpatrick, Liu, Mainzer, Marsh, Masci,
  McCallon, Padgett, Ressler, Royer, Skrutskie, Stanford, Wyatt, Tholen, Tsai,
  Wachter, Wheelock, Yan, Alles, Beck, Grav, Masiero, McCollum, McGehee, \&
  Wittman}]{cutri_explanatory_2011}
Cutri, R.~M., Wright, E.~L., Conrow, T., {et~al.} 2011, Explanatory
  {Supplement} to the {WISE} {Preliminary} {Data} {Release} {Products}, Tech.
  rep.

\bibitem[{Endres {et~al.}(2016)Endres, Schlemmer, Schilke, Stutzki, \&
  Müller}]{endres_cologne_2016}
Endres, C.~P., Schlemmer, S., Schilke, P., Stutzki, J., \& Müller, H. S.~P.
  2016, Journal of Molecular Spectroscopy, 327, 95

\bibitem[{Fehér {et~al.}(2017)Fehér, Juvela, Lunttila, Montillaud,
  Ristorcelli, Zahorecz, \& Tóth}]{feher_co_2017}
Fehér, O., Juvela, M., Lunttila, T., {et~al.} 2017, A\&A, 606, A102

\bibitem[{Friesen {et~al.}(2017)Friesen, Pineda, co~PIs, Rosolowsky, Alves,
  Chacón-Tanarro, How-Huan~Chen, Chun-Yuan~Chen, Di~Francesco, Keown, Kirk,
  Punanova, Seo, Shirley, Ginsburg, Hall, Offner, Singh, Arce, Caselli,
  Goodman, Martin, Matzner, Myers, Redaelli, \&
  Collaboration}]{friesen_green_2017}
Friesen, R.~K., Pineda, J.~E., co~PIs, {et~al.} 2017, ApJ, 843, 63

\bibitem[{{Gaia Collaboration} {et~al.}(2016){Gaia Collaboration}, Prusti,
  de~Bruijne, Brown, Vallenari, Babusiaux, Bailer-Jones, Bastian, Biermann,
  Evans, Eyer, Jansen, Jordi, Klioner, Lammers, Lindegren, Luri, Mignard,
  Milligan, Panem, Poinsignon, Pourbaix, Randich, Sarri, Sartoretti, Siddiqui,
  Soubiran, Valette, van Leeuwen, Walton, Aerts, Arenou, Cropper, Drimmel,
  Høg, Katz, Lattanzi, O'Mullane, Grebel, Holland, Huc, Passot, Bramante,
  Cacciari, Castañeda, Chaoul, Cheek, De~Angeli, Fabricius, Guerra,
  Hernández, Jean-Antoine-Piccolo, Masana, Messineo, Mowlavi, Nienartowicz,
  Ordóñez-Blanco, Panuzzo, Portell, Richards, Riello, Seabroke, Tanga,
  Thévenin, Torra, Els, Gracia-Abril, Comoretto, Garcia-Reinaldos, Lock,
  Mercier, Altmann, Andrae, Astraatmadja, Bellas-Velidis, Benson, Berthier,
  Blomme, Busso, Carry, Cellino, Clementini, Cowell, Creevey, Cuypers,
  Davidson, De~Ridder, de~Torres, Delchambre, Dell'Oro, Ducourant, Frémat,
  García-Torres, Gosset, Halbwachs, Hambly, Harrison, Hauser, Hestroffer,
  Hodgkin, Huckle, Hutton, Jasniewicz, Jordan, Kontizas, Korn, Lanzafame,
  Manteiga, Moitinho, Muinonen, Osinde, Pancino, Pauwels, Petit, Recio-Blanco,
  Robin, Sarro, Siopis, Smith, Smith, Sozzetti, Thuillot, van Reeven, Viala,
  Abbas, Abreu~Aramburu, Accart, Aguado, Allan, Allasia, Altavilla, Álvarez,
  Alves, Anderson, Andrei, Anglada~Varela, Antiche, Antoja, Antón, Arcay,
  Atzei, Ayache, Bach, Baker, Balaguer-Núñez, Barache, Barata, Barbier,
  Barblan, Baroni, Barrado~y Navascués, Barros, Barstow, Becciani, Bellazzini,
  Bellei, Bello~García, Belokurov, Bendjoya, Berihuete, Bianchi, Bienaymé,
  Billebaud, Blagorodnova, Blanco-Cuaresma, Boch, Bombrun, Borrachero,
  Bouquillon, Bourda, Bouy, Bragaglia, Breddels, Brouillet, Brüsemeister,
  Bucciarelli, Budnik, Burgess, Burgon, Burlacu, Busonero, Buzzi, Caffau,
  Cambras, Campbell, Cancelliere, Cantat-Gaudin, Carlucci, Carrasco,
  Castellani, Charlot, Charnas, Charvet, Chassat, Chiavassa, Clotet, Cocozza,
  Collins, Collins, Costigan, Crifo, Cross, Crosta, Crowley, Dafonte, Damerdji,
  Dapergolas, David, David, De~Cat, de~Felice, de~Laverny, De~Luise, De~March,
  de~Martino, de~Souza, Debosscher, del Pozo, Delbo, Delgado, Delgado,
  di~Marco, Di~Matteo, Diakite, Distefano, Dolding, Dos~Anjos, Drazinos,
  Durán, Dzigan, Ecale, Edvardsson, Enke, Erdmann, Escolar, Espina, Evans,
  Eynard~Bontemps, Fabre, Fabrizio, Faigler, Falcão, Farràs~Casas, Faye,
  Federici, Fedorets, Fernández-Hernández, Fernique, Fienga, Figueras,
  Filippi, Findeisen, Fonti, Fouesneau, Fraile, Fraser, Fuchs, Furnell, Gai,
  Galleti, Galluccio, Garabato, García-Sedano, Garé, Garofalo, Garralda,
  Gavras, Gerssen, Geyer, Gilmore, Girona, Giuffrida, Gomes, González-Marcos,
  González-Núñez, González-Vidal, Granvik, Guerrier, Guillout, Guiraud,
  Gúrpide, Gutiérrez-Sánchez, Guy, Haigron, Hatzidimitriou, Haywood, Heiter,
  Helmi, Hobbs, Hofmann, Holl, Holland, Hunt, Hypki, Icardi, Irwin, Jevardat~de
  Fombelle, Jofré, Jonker, Jorissen, Julbe, Karampelas, Kochoska, Kohley,
  Kolenberg, Kontizas, Koposov, Kordopatis, Koubsky, Kowalczyk, Krone-Martins,
  Kudryashova, Kull, Bachchan, Lacoste-Seris, Lanza, Lavigne,
  Le~Poncin-Lafitte, Lebreton, Lebzelter, Leccia, Leclerc, Lecoeur-Taibi,
  Lemaitre, Lenhardt, Leroux, Liao, Licata, Lindstrøm, Lister, Livanou, Lobel,
  Löffler, López, Lopez-Lozano, Lorenz, Loureiro, MacDonald,
  Magalhães~Fernandes, Managau, Mann, Mantelet, Marchal, Marchant, Marconi,
  Marie, Marinoni, Marrese, Marschalkó, Marshall, Martín-Fleitas, Martino,
  Mary, Matijevič, Mazeh, McMillan, Messina, Mestre, Michalik, Millar,
  Miranda, Molina, Molinaro, Molinaro, Molnár, Moniez, Montegriffo, Monteiro,
  Mor, Mora, Morbidelli, Morel, Morgenthaler, Morley, Morris, Mulone, Muraveva,
  Musella, Narbonne, Nelemans, Nicastro, Noval, Ordénovic, Ordieres-Meré,
  Osborne, Pagani, Pagano, Pailler, Palacin, Palaversa, Parsons, Paulsen,
  Pecoraro, Pedrosa, Pentikäinen, Pereira, Pichon, Piersimoni, Pineau, Plachy,
  Plum, Poujoulet, Prša, Pulone, Ragaini, Rago, Rambaux, Ramos-Lerate,
  Ranalli, Rauw, Read, Regibo, Renk, Reylé, Ribeiro, Rimoldini, Ripepi, Riva,
  Rixon, Roelens, Romero-Gómez, Rowell, Royer, Rudolph, Ruiz-Dern, Sadowski,
  Sagristà~Sellés, Sahlmann, Salgado, Salguero, Sarasso, Savietto, Schnorhk,
  Schultheis, Sciacca, Segol, Segovia, Segransan, Serpell, Shih, Smareglia,
  Smart, Smith, Solano, Solitro, Sordo, Soria~Nieto, Souchay, Spagna, Spoto,
  Stampa, Steele, Steidelmüller, Stephenson, Stoev, Suess, Süveges, Surdej,
  Szabados, Szegedi-Elek, Tapiador, Taris, Tauran, Taylor, Teixeira, Terrett,
  Tingley, Trager, Turon, Ulla, Utrilla, Valentini, van Elteren, Van~Hemelryck,
  van Leeuwen, Varadi, Vecchiato, Veljanoski, Via, Vicente, Vogt, Voss,
  Votruba, Voutsinas, Walmsley, Weiler, Weingrill, Werner, Wevers, Whitehead,
  Wyrzykowski, Yoldas, Žerjal, Zucker, Zurbach, Zwitter, Alecu, Allen,
  Allende~Prieto, Amorim, Anglada-Escudé, Arsenijevic, Azaz, Balm, Beck,
  Bernstein, Bigot, Bijaoui, Blasco, Bonfigli, Bono, Boudreault, Bressan,
  Brown, Brunet, Bunclark, Buonanno, Butkevich, Carret, Carrion, Chemin,
  Chéreau, Corcione, Darmigny, de~Boer, de~Teodoro, de~Zeeuw, Delle~Luche,
  Domingues, Dubath, Fodor, Frézouls, Fries, Fustes, Fyfe, Gallardo, Gallegos,
  Gardiol, Gebran, Gomboc, Gómez, Grux, Gueguen, Heyrovsky, Hoar, Iannicola,
  Isasi~Parache, Janotto, Joliet, Jonckheere, Keil, Kim, Klagyivik, Klar,
  Knude, Kochukhov, Kolka, Kos, Kutka, Lainey, LeBouquin, Liu, Loreggia,
  Makarov, Marseille, Martayan, Martinez-Rubi, Massart, Meynadier, Mignot,
  Munari, Nguyen, Nordlander, Ocvirk, O'Flaherty, Olias~Sanz, Ortiz, Osorio,
  Oszkiewicz, Ouzounis, Palmer, Park, Pasquato, Peltzer, Peralta, Péturaud,
  Pieniluoma, Pigozzi, Poels, Prat, Prod'homme, Raison, Rebordao, Risquez,
  Rocca-Volmerange, Rosen, Ruiz-Fuertes, Russo, Sembay, Serraller~Vizcaino,
  Short, Siebert, Silva, Sinachopoulos, Slezak, Soffel, Sosnowska, Straižys,
  ter Linden, Terrell, Theil, Tiede, Troisi, Tsalmantza, Tur, Vaccari, Vachier,
  Valles, Van~Hamme, Veltz, Virtanen, Wallut, Wichmann, Wilkinson, Ziaeepour,
  \& Zschocke}]{gaia_collaboration_gaia_2016}
{Gaia Collaboration}, Prusti, T., de~Bruijne, J. H.~J., {et~al.} 2016,
  Astronomy and Astrophysics, 595, A1

\bibitem[{Gratier {et~al.}(2017)Gratier, Bron, Gerin, Pety, Guzman, Orkisz,
  Bardeau, Goicoechea, Le~Petit, Liszt, Öberg, Peretto, Roueff, Sievers, \&
  Tremblin}]{gratier_dissecting_2017}
Gratier, P., Bron, E., Gerin, M., {et~al.} 2017, A\&A, 599, A100

\bibitem[{Griffin {et~al.}(2010)Griffin, Abergel, Abreu, Ade, André, Augueres,
  Babbedge, Bae, Baillie, Baluteau, Barlow, Bendo, Benielli, Bock, Bonhomme,
  Brisbin, Brockley-Blatt, Caldwell, Cara, Castro-Rodriguez, Cerulli, Chanial,
  Chen, Clark, Clements, Clerc, Coker, Communal, Conversi, Cox, Crumb,
  Cunningham, Daly, Davis, de~Antoni, Delderfield, Devin, di~Giorgio,
  Didschuns, Dohlen, Donati, Dowell, Dowell, Duband, Dumaye, Emery, Ferlet,
  Ferrand, Fontignie, Fox, Franceschini, Frerking, Fulton, Garcia, Gastaud,
  Gear, Glenn, Goizel, Griffin, Grundy, Guest, Guillemet, Hargrave, Harwit,
  Hastings, Hatziminaoglou, Herman, Hinde, Hristov, Huang, Imhof, Isaak,
  Israelsson, Ivison, Jennings, Kiernan, King, Lange, Latter, Laurent, Laurent,
  Leeks, Lellouch, Levenson, Li, Li, Lilienthal, Lim, Liu, Lu, Madden,
  Mainetti, Marliani, McKay, Mercier, Molinari, Morris, Moseley, Mulder, Mur,
  Naylor, Nguyen, O'Halloran, Oliver, Olofsson, Olofsson, Orfei, Page, Pain,
  Panuzzo, Papageorgiou, Parks, Parr-Burman, Pearce, Pearson, Pérez-Fournon,
  Pinsard, Pisano, Podosek, Pohlen, Polehampton, Pouliquen, Rigopoulou, Rizzo,
  Roseboom, Roussel, Rowan-Robinson, Rownd, Saraceno, Sauvage, Savage, Savini,
  Sawyer, Scharmberg, Schmitt, Schneider, Schulz, Schwartz, Shafer, Shupe,
  Sibthorpe, Sidher, Smith, Smith, Smith, Spencer, Stobie, Sudiwala, Sukhatme,
  Surace, Stevens, Swinyard, Trichas, Tourette, Triou, Tseng, Tucker, Turner,
  Vaccari, Valtchanov, Vigroux, Virique, Voellmer, Walker, Ward, Waskett,
  Weilert, Wesson, White, Whitehouse, Wilson, Winter, Woodcraft, Wright, Xu,
  Zavagno, Zemcov, Zhang, \& Zonca}]{griffin_herschel-spire_2010}
Griffin, M.~J., Abergel, A., Abreu, A., {et~al.} 2010, A\&A, 518, L3

\bibitem[{Juvela {et~al.}(2010)Juvela, Ristorcelli, Montier, Marshall,
  Pelkonen, Malinen, Ysard, Tóth, Harju, Bernard, Schneider, Verebélyi,
  Anderson, André, Giard, Krause, Lehtinen, Macias-Perez, Martin, McGehee,
  Meny, Motte, Pagani, Paladini, Reach, Valenziano, Ward-Thompson, \&
  Zavagno}]{juvela_galactic_2010}
Juvela, M., Ristorcelli, I., Montier, L.~A., {et~al.} 2010, A\&A, 518, L93

\bibitem[{Juvela {et~al.}(2012)Juvela, Ristorcelli, Pagani, Doi, Pelkonen,
  Marshall, Bernard, Falgarone, Malinen, Marton, McGehee, Montier, Motte,
  Paladini, Tóth, Ysard, Zahorecz, \& Zavagno}]{juvela_galactic_2012}
Juvela, M., Ristorcelli, I., Pagani, L., {et~al.} 2012, A\&A, 541, 12

\bibitem[{MacLaren {et~al.}(1988)MacLaren, Richardson, \&
  Wolfendale}]{maclaren_corrections_1988}
MacLaren, I., Richardson, K.~M., \& Wolfendale, A.~W. 1988, ApJ, 333, 821

\bibitem[{Marton {et~al.}(2016)Marton, Tóth, Paladini, Kun, Zahorecz, McGehee,
  \& Kiss}]{marton_all-sky_2016}
Marton, G., Tóth, L.~V., Paladini, R., {et~al.} 2016, MNRAS, 458, 3479

\bibitem[{Montier {et~al.}(2010)Montier, Pelkonen, Juvela, Ristorcelli, \&
  Marshall}]{montier_all-sky_2010}
Montier, L.~A., Pelkonen, V.-M., Juvela, M., Ristorcelli, I., \& Marshall,
  D.~J. 2010, A\&A, 522, 83

\bibitem[{Montillaud {et~al.}(2015)Montillaud, Juvela, Rivera-Ingraham,
  Malinen, Pelkonen, Ristorcelli, Montier, Marshall, Marton, Pagani, Toth,
  Zahorecz, Ysard, McGehee, Paladini, Falgarone, Bernard, Motte, Zavagno, \&
  Doi}]{montillaud_galactic_2015}
Montillaud, J., Juvela, M., Rivera-Ingraham, A., {et~al.} 2015, A\&A, 584, A92

\bibitem[{Orkisz {et~al.}(2019)Orkisz, Peretto, Pety, Gerin, Levrier, Bron,
  Bardeau, Goicoechea, Gratier, Guzmán, Hughes, Languignon, Le~Petit, Liszt,
  Öberg, Roueff, Sievers, \& Tremblin}]{orkisz_dynamically_2019}
Orkisz, J.~H., Peretto, N., Pety, J., {et~al.} 2019, A\&A, 624, A113

\bibitem[{Orkisz {et~al.}(2017)Orkisz, Pety, Gerin, Bron, Guzmán, Bardeau,
  Goicoechea, Gratier, Le~Petit, Levrier, Liszt, Öberg, Peretto, Roueff,
  Sievers, \& Tremblin}]{orkisz_turbulence_2017}
Orkisz, J.~H., Pety, J., Gerin, M., {et~al.} 2017, A\&A, 599, A99

\bibitem[{Padoan {et~al.}(2017)Padoan, Haugbølle, Nordlund, \&
  Frimann}]{padoan_supernova_2017}
Padoan, P., Haugbølle, T., Nordlund, A., \& Frimann, S. 2017, ApJ, 840, 48

\bibitem[{Pagani {et~al.}(2009)Pagani, Daniel, \&
  Dubernet}]{pagani_frequency_2009}
Pagani, L., Daniel, F., \& Dubernet, M.-L. 2009, A\&A, 494, 719

\bibitem[{Pety {et~al.}(2017)Pety, Guzmán, Orkisz, Liszt, Gerin, Bron,
  Bardeau, Goicoechea, Gratier, Le~Petit, Levrier, Öberg, Roueff, \&
  Sievers}]{pety_anatomy_2017}
Pety, J., Guzmán, V.~V., Orkisz, J.~H., {et~al.} 2017, A\&A, 599, A98

\bibitem[{Pineda {et~al.}(2010)Pineda, Goldsmith, Chapman, Snell, Li,
  Cambrésy, \& Brunt}]{pineda_relation_2010}
Pineda, J.~L., Goldsmith, P.~F., Chapman, N., {et~al.} 2010, ApJ, 721, 686

\bibitem[{Planck~Collaboration {et~al.}(2011)Planck~Collaboration, Ade,
  Aghanim, Arnaud, Ashdown, Aumont, Baccigalupi, Balbi, Banday, Barreiro,
  Bartlett, Battaner, Benabed, Benoît, Bernard, Bersanelli, Bhatia, Bock,
  Bonaldi, Bond, Borrill, Bouchet, Boulanger, Bucher, Burigana, Cabella,
  Cantalupo, Cardoso, Catalano, Cayón, Challinor, Chamballu, Chary, Chiang,
  Christensen, Clements, Colombi, Couchot, Coulais, Crill, Cuttaia, Danese,
  Davies, Davis, de~Bernardis, de~Gasperis, de~Rosa, de~Zotti, Delabrouille,
  Delouis, Désert, Dickinson, Dobashi, Donzelli, Doré, Dörl, Douspis, Dupac,
  Efstathiou, Enßlin, Falgarone, Finelli, Forni, Frailis, Franceschi,
  Galeotta, Ganga, Giard, Giardino, Giraud-Héraud, González-Nuevo, Górski,
  Gratton, Gregorio, Gruppuso, Hansen, Harrison, Helou, Henrot-Versillé,
  Herranz, Hildebrandt, Hivon, Hobson, Holmes, Hovest, Hoyland, Huffenberger,
  Jaffe, Joncas, Jones, Juvela, Keihänen, Keskitalo, Kisner, Kneissl, Knox,
  Kurki-Suonio, Lagache, Lamarre, Lasenby, Laureijs, Lawrence, Leach, Leonardi,
  Leroy, Linden-Vørnle, López-Caniego, Lubin, Macías-Pérez, MacTavish,
  Maffei, Mandolesi, Mann, Maris, Marshall, Martin, Martínez-González,
  Marton, Masi, Matarrese, Matthai, Mazzotta, McGehee, Melchiorri, Mendes,
  Mennella, Mitra, Miville-Deschênes, Moneti, Montier, Morgante, Mortlock,
  Munshi, Murphy, Naselsky, Nati, Natoli, Netterfield, Nørgaard-Nielsen,
  Noviello, Novikov, Novikov, Osborne, Pajot, Paladini, Pasian, Patanchon,
  Pearson, Pelkonen, Perdereau, Perotto, Perrotta, Piacentini, Piat,
  Plaszczynski, Pointecouteau, Polenta, Ponthieu, Poutanen, Prézeau, Prunet,
  Puget, Reach, Rebolo, Reinecke, Renault, Ricciardi, Riller, Ristorcelli,
  Rocha, Rosset, Rowan-Robinson, Rubiño-Martín, Rusholme, Sandri, Santos,
  Savini, Scott, Seiffert, Smoot, Starck, Stivoli, Stolyarov, Sudiwala, Sygnet,
  Tauber, Terenzi, Toffolatti, Tomasi, Torre, Toth, Tristram, Tuovinen, Umana,
  Valenziano, Vielva, Villa, Vittorio, Wade, Wandelt, Ysard, Yvon, Zacchei,
  Zahorecz, \& Zonca}]{planck_collaboration_planck_2011}
Planck~Collaboration, ., Ade, P. A.~R., Aghanim, N., {et~al.} 2011, A\&A, 536,
  23

\bibitem[{{Planck Collaboration} {et~al.}(2016){Planck Collaboration}, Ade,
  Aghanim, Arnaud, Ashdown, Aumont, Baccigalupi, Banday, Barreiro, Bartolo,
  Battaner, Benabed, Benoît, Benoit-Lévy, Bernard, Bersanelli, Bielewicz,
  Bonaldi, Bonavera, Bond, Borrill, Bouchet, Boulanger, Bucher, Burigana,
  Butler, Calabrese, Catalano, Chamballu, Chiang, Christensen, Clements,
  Colombi, Colombo, Combet, Couchot, Coulais, Crill, Curto, Cuttaia, Danese,
  Davies, Davis, de~Bernardis, de~Rosa, de~Zotti, Delabrouille, Désert,
  Dickinson, Diego, Dole, Donzelli, Doré, Douspis, Ducout, Dupac, Efstathiou,
  Elsner, Enßlin, Eriksen, Falgarone, Fergusson, Finelli, Forni, Frailis,
  Fraisse, Franceschi, Frejsel, Galeotta, Galli, Ganga, Giard, Giraud-Héraud,
  Gjerløw, González-Nuevo, Górski, Gratton, Gregorio, Gruppuso, Gudmundsson,
  Hansen, Hanson, Harrison, Helou, Henrot-Versillé, Hernández-Monteagudo,
  Herranz, Hildebrandt, Hivon, Hobson, Holmes, Hornstrup, Hovest, Huffenberger,
  Hurier, Jaffe, Jaffe, Jones, Juvela, Keihänen, Keskitalo, Kisner, Knoche,
  Kunz, Kurki-Suonio, Lagache, Lamarre, Lasenby, Lattanzi, Lawrence, Leonardi,
  Lesgourgues, Levrier, Liguori, Lilje, Linden-Vørnle, López-Caniego, Lubin,
  Macías-Pérez, Maggio, Maino, Mandolesi, Mangilli, Marshall, Martin,
  Martínez-González, Masi, Matarrese, Mazzotta, McGehee, Melchiorri, Mendes,
  Mennella, Migliaccio, Mitra, Miville-Deschênes, Moneti, Montier, Morgante,
  Mortlock, Moss, Munshi, Murphy, Naselsky, Nati, Natoli, Netterfield,
  Nørgaard-Nielsen, Noviello, Novikov, Novikov, Oxborrow, Paci, Pagano, Pajot,
  Paladini, Paoletti, Pasian, Patanchon, Pearson, Pelkonen, Perdereau, Perotto,
  Perrotta, Pettorino, Piacentini, Piat, Pierpaoli, Pietrobon, Plaszczynski,
  Pointecouteau, Polenta, Pratt, Prézeau, Prunet, Puget, Rachen, Reach,
  Rebolo, Reinecke, Remazeilles, Renault, Renzi, Ristorcelli, Rocha, Rosset,
  Rossetti, Roudier, Rubiño-Martín, Rusholme, Sandri, Santos, Savelainen,
  Savini, Scott, Seiffert, Shellard, Spencer, Stolyarov, Sudiwala, Sunyaev,
  Sutton, Suur-Uski, Sygnet, Tauber, Terenzi, Toffolatti, Tomasi, Tristram,
  Tucci, Tuovinen, Umana, Valenziano, Valiviita, Van~Tent, Vielva, Villa, Wade,
  Wandelt, Wehus, Yvon, Zacchei, \& Zonca}]{planck_collaboration_planck_2016}
{Planck Collaboration}, Ade, P. A.~R., Aghanim, N., {et~al.} 2016, A\&A, 594,
  A28

\bibitem[{Poglitsch {et~al.}(2010)Poglitsch, Waelkens, Geis, Feuchtgruber,
  Vandenbussche, Rodriguez, Krause, Renotte, van Hoof, Saraceno, Cepa,
  Kerschbaum, Agnèse, Ali, Altieri, Andreani, Augueres, Balog, Barl, Bauer,
  Belbachir, Benedettini, Billot, Boulade, Bischof, Blommaert, Callut, Cara,
  Cerulli, Cesarsky, Contursi, Creten, De~Meester, Doublier, Doumayrou, Duband,
  Exter, Genzel, Gillis, Grözinger, Henning, Herreros, Huygen, Inguscio,
  Jakob, Jamar, Jean, de~Jong, Katterloher, Kiss, Klaas, Lemke, Lutz, Madden,
  Marquet, Martignac, Mazy, Merken, Montfort, Morbidelli, Müller, Nielbock,
  Okumura, Orfei, Ottensamer, Pezzuto, Popesso, Putzeys, Regibo, Reveret,
  Royer, Sauvage, Schreiber, Stegmaier, Schmitt, Schubert, Sturm, Thiel,
  Tofani, Vavrek, Wetzstein, Wieprecht, \&
  Wiezorrek}]{poglitsch_photodetector_2010}
Poglitsch, A., Waelkens, C., Geis, N., {et~al.} 2010, A\&A, 518, L2

\bibitem[{Rapson {et~al.}(2014)Rapson, Pipher, Gutermuth, Megeath, Allen,
  Myers, \& Allen}]{rapson_spitzer_2014}
Rapson, V.~A., Pipher, J.~L., Gutermuth, R.~A., {et~al.} 2014, ApJ, 794, 124

\bibitem[{Renaud {et~al.}(2013)Renaud, Bournaud, Emsellem, Elmegreen, Teyssier,
  Alves, Chapon, Combes, Dekel, Gabor, Hennebelle, \&
  Kraljic}]{renaud_sub-parsec_2013}
Renaud, F., Bournaud, F., Emsellem, E., {et~al.} 2013, MNRAS, 436, 1836

\bibitem[{Sanhueza {et~al.}(2012)Sanhueza, Jackson, Foster, Garay, Silva, \&
  Finn}]{sanhueza_chemistry_2012}
Sanhueza, P., Jackson, J.~M., Foster, J.~B., {et~al.} 2012, ApJ, 756, 60

\bibitem[{Venuti {et~al.}(2018)Venuti, Prisinzano, Sacco, Flaccomio, Bonito,
  Damiani, Micela, Guarcello, Randich, Stauffer, Cody, Jeffries, Alencar,
  Alfaro, Lanzafame, Pancino, Bayo, Carraro, Costado, Frasca, Jofré,
  Morbidelli, Sousa, \& Zaggia}]{venuti_gaia-eso_2018}
Venuti, L., Prisinzano, L., Sacco, G.~G., {et~al.} 2018, A\&A, 609, A10

\bibitem[{Vázquez-Semadeni {et~al.}(2009)Vázquez-Semadeni, Gómez, Jappsen,
  Ballesteros-Paredes, \& Klessen}]{vazquez-semadeni_high-_2009}
Vázquez-Semadeni, E., Gómez, G.~C., Jappsen, A.-K., Ballesteros-Paredes, J.,
  \& Klessen, R.~S. 2009, ApJ, 707, 1023

\bibitem[{Vázquez-Semadeni {et~al.}(2019)Vázquez-Semadeni, Palau,
  Ballesteros-Paredes, Gómez, \&
  Zamora-Avilés}]{vazquez-semadeni_global_2019}
Vázquez-Semadeni, E., Palau, A., Ballesteros-Paredes, J., Gómez, G.~C., \&
  Zamora-Avilés, M. 2019, arXiv e-prints, arXiv:1903.11247

\bibitem[{Williams(2018)}]{williams_evolution_2018}
Williams, G.~M. 2018, Ph.D. Thesis

\bibitem[{Wilson {et~al.}(2013)Wilson, Rohlfs, \&
  Hüttemeister}]{wilson_tools_2013}
Wilson, T.~L., Rohlfs, K., \& Hüttemeister, S. 2013, Tools of {Radio}
  {Astronomy} (Springer Science \& Business Media), google-Books-ID:
  DoDBBAAAQBAJ

\bibitem[{Wilson \& Rood(1994)}]{wilson_abundances_1994}
Wilson, T.~L. \& Rood, R. 1994, ARA\&A, 32, 191

\bibitem[{Wolf-Chase {et~al.}(2003)Wolf-Chase, Moriarty-Schieven, Fich, \&
  Barsony}]{wolf-chase_star_2003}
Wolf-Chase, G., Moriarty-Schieven, G., Fich, M., \& Barsony, M. 2003, MNRAS,
  344, 809

\bibitem[{Wright {et~al.}(2010)Wright, Eisenhardt, Mainzer, Ressler, Cutri,
  Jarrett, Kirkpatrick, Padgett, McMillan, Skrutskie, Stanford, Cohen, Walker,
  Mather, Leisawitz, Gautier, McLean, Benford, Lonsdale, Blain, Mendez, Irace,
  Duval, Liu, Royer, Heinrichsen, Howard, Shannon, Kendall, Walsh, Larsen,
  Cardon, Schick, Schwalm, Abid, Fabinsky, Naes, \&
  Tsai}]{wright_wide-field_2010}
Wright, E.~L., Eisenhardt, P. R.~M., Mainzer, A.~K., {et~al.} 2010, AJ, 140,
  1868

\bibitem[{Zhao {et~al.}(2015)Zhao, Galazutdinov, Linnartz, \&
  Krełowski}]{zhao_mercapto_2015}
Zhao, D., Galazutdinov, G.~A., Linnartz, H., \& Krełowski, J. 2015, A\&A, 579,
  L1

\end{thebibliography}

\appendix

\section{Analysis of molecular lines}

   \subsection{Examples of combined line fitting}\label{sec:fit_examples}

\begin{figure*}
      \includegraphics[width=0.49\textwidth]{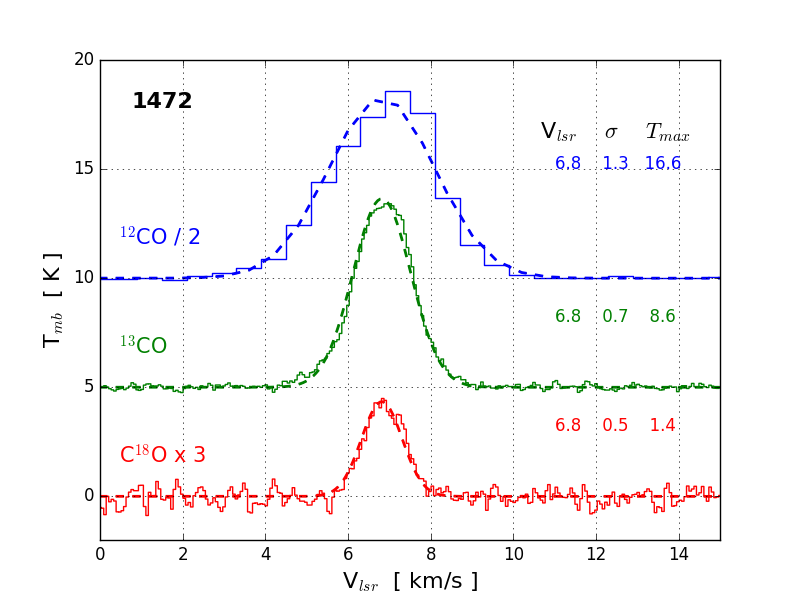}
      \includegraphics[width=0.49\textwidth]{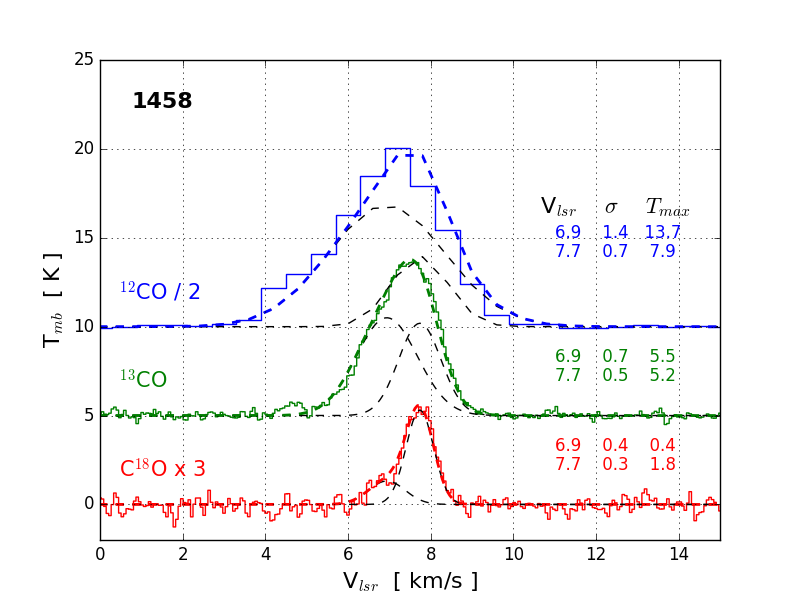}\\
      \includegraphics[width=0.49\textwidth]{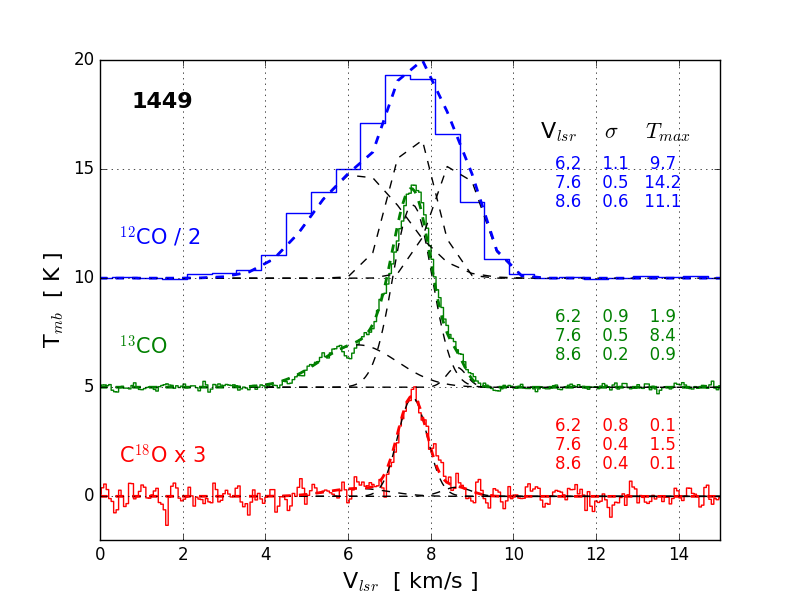}
      \includegraphics[width=0.49\textwidth]{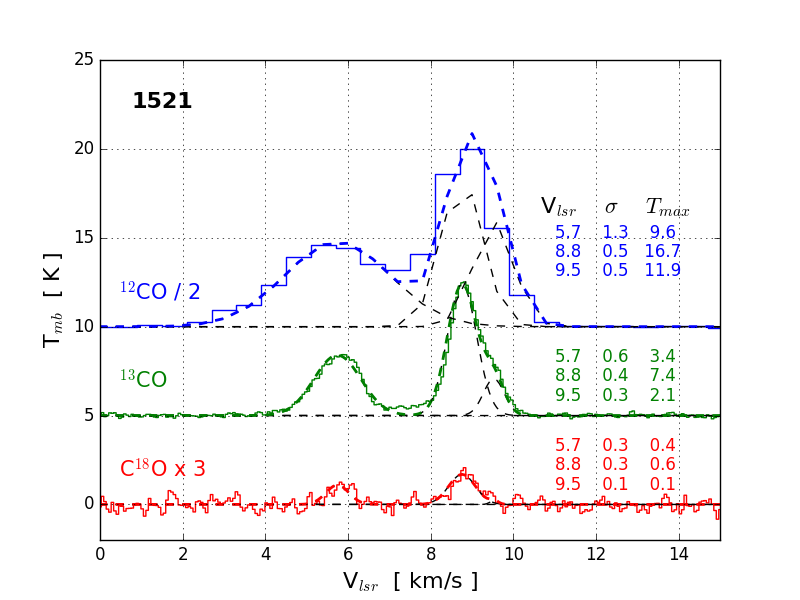}
      \caption{Examples of combined line fitting with locked velocities for sources 
      s1472, s1458, s1449, and s1521, as indicated in the top-left corner of each 
      frame. The best-fit values are also given in the frames in units of km\,s$^{-1}$, km\,s$^{-1}$, 
      and K for the central velocity \Vlsr, the velocity dispersion $\sigma$, and 
      the peak main beam temperature \Tmb, respectively.}
      \label{fig:fit_examples}
   \end{figure*}

   As presented in Sect.~\ref{sec:method}, we performed a combined line fitting 
   of \CO, \tCO, and \CeO\, data where the same number of velocity components
   was used for the three lines, and the central velocity of each component was
   locked between the three transitions. The fitting procedure was semi-automatic
   but the small number of sources enabled us to check each fit by eye.
   Figure~\ref{fig:fit_examples} shows a few examples of typical cases with one
   Gaussian component (s1472), two Gaussian components (s1458) and three Gaussian 
   components (s1521 and s1449).

\begin{figure*}
      \includegraphics[width=0.33\textwidth]{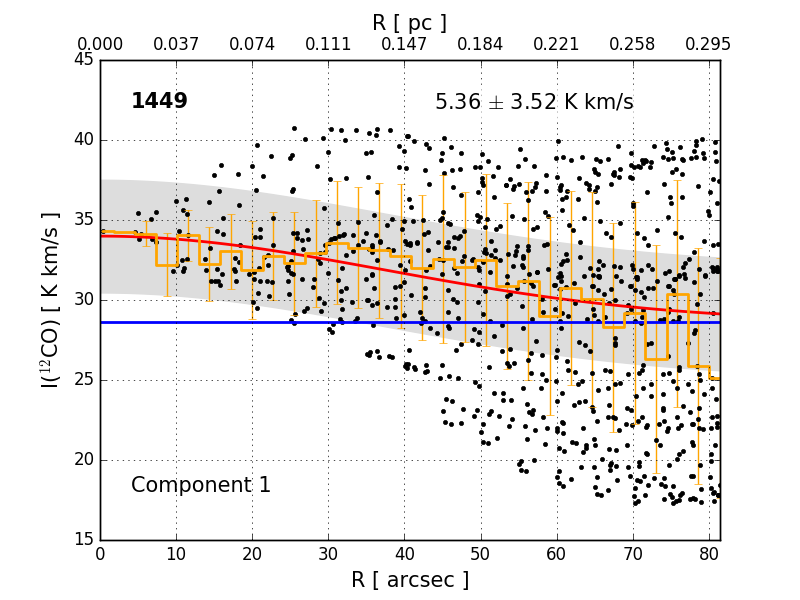}
      \includegraphics[width=0.33\textwidth]{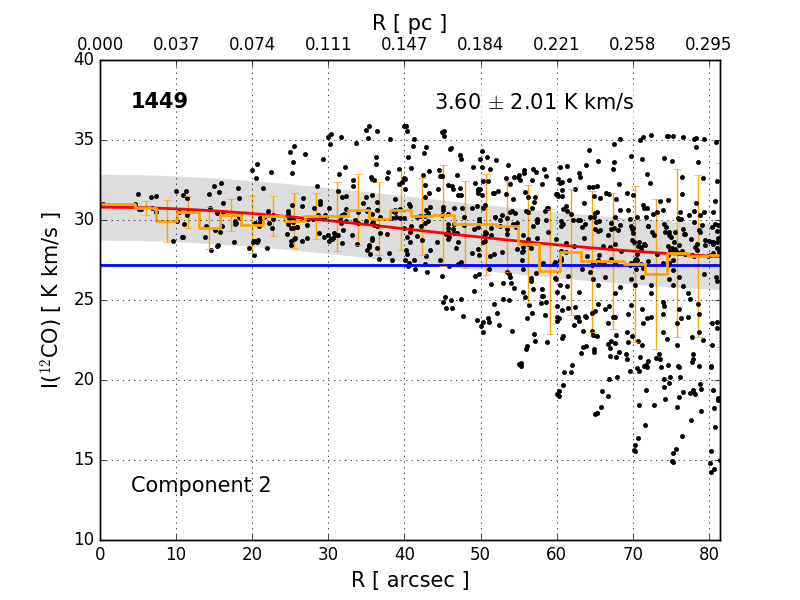}
      \includegraphics[width=0.33\textwidth]{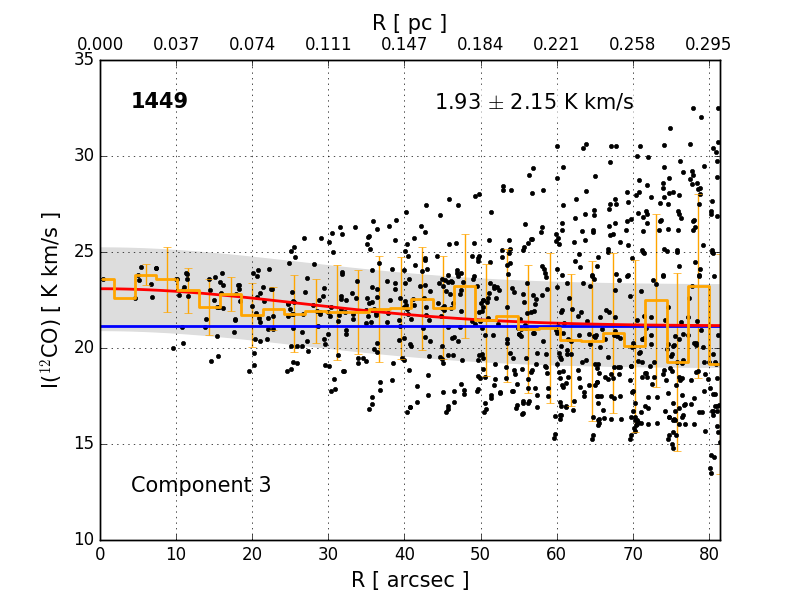}\\
      \includegraphics[width=0.33\textwidth]{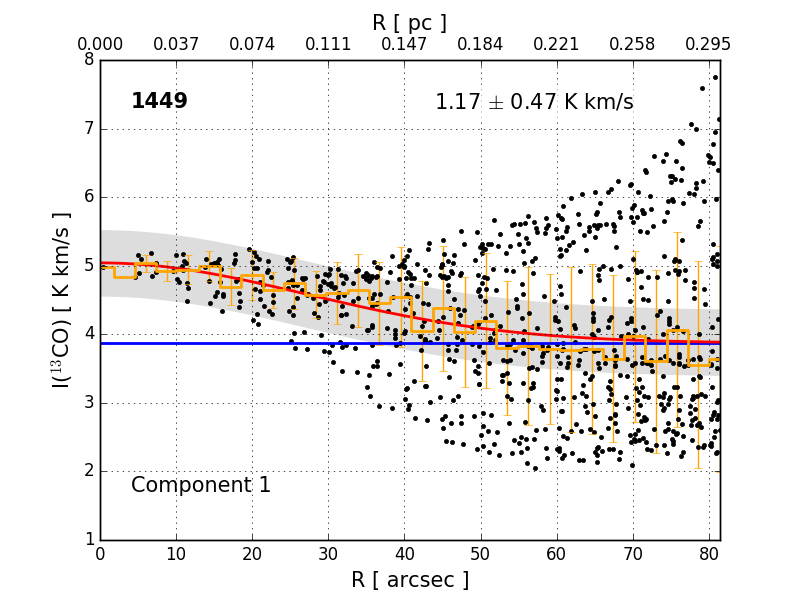}
      \includegraphics[width=0.33\textwidth]{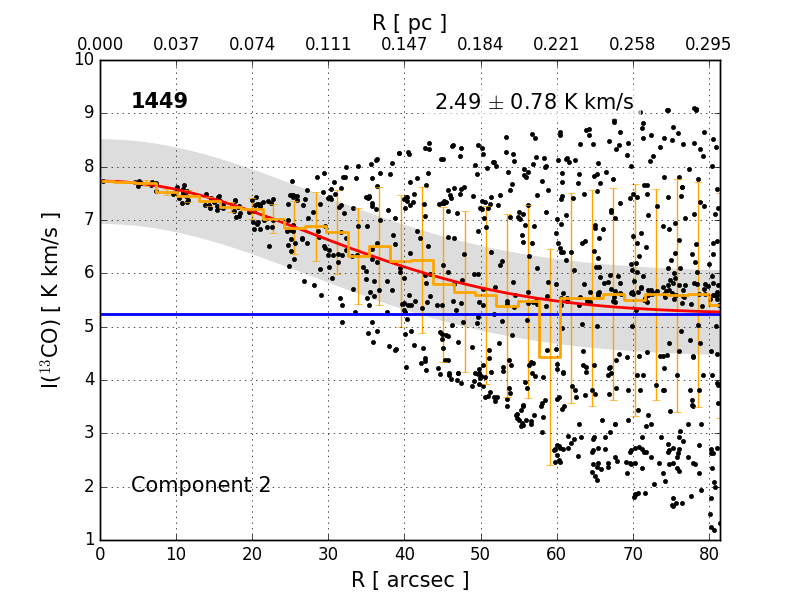}
      \includegraphics[width=0.33\textwidth]{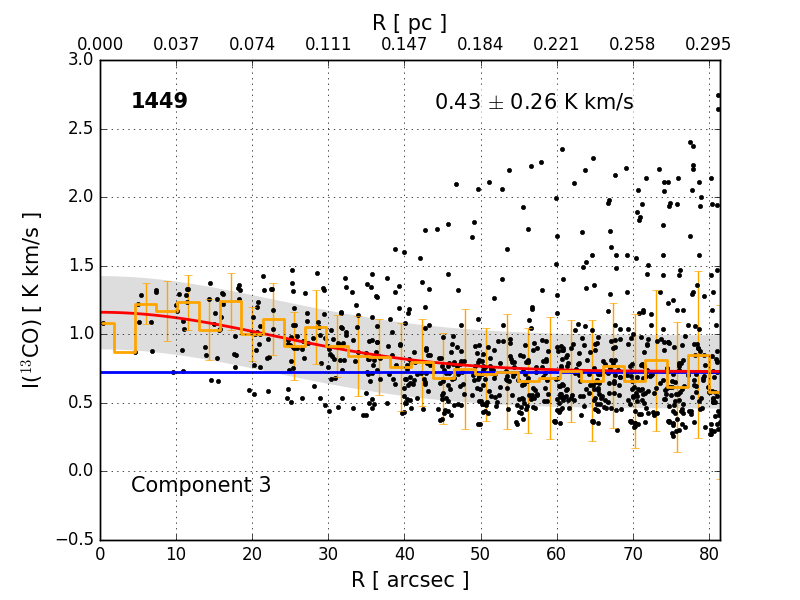}\\
      \includegraphics[width=0.33\textwidth]{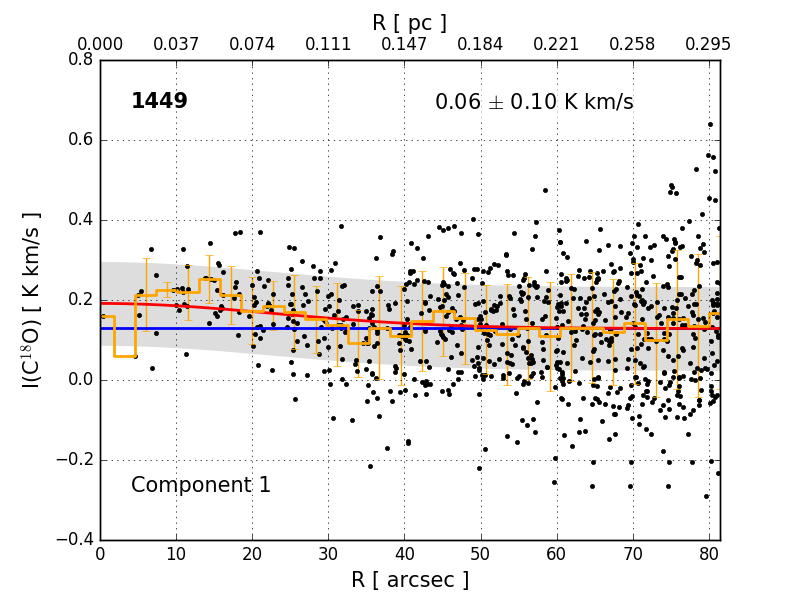}
      \includegraphics[width=0.33\textwidth]{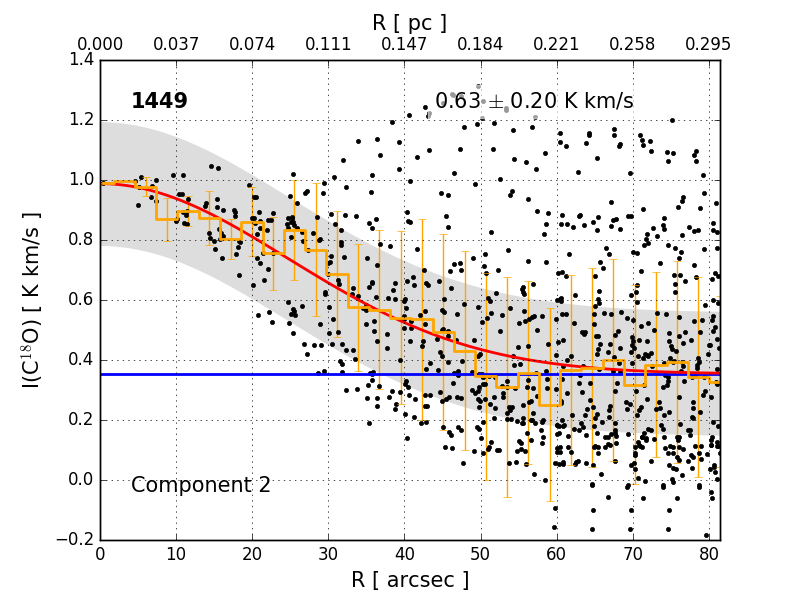}
      \includegraphics[width=0.33\textwidth]{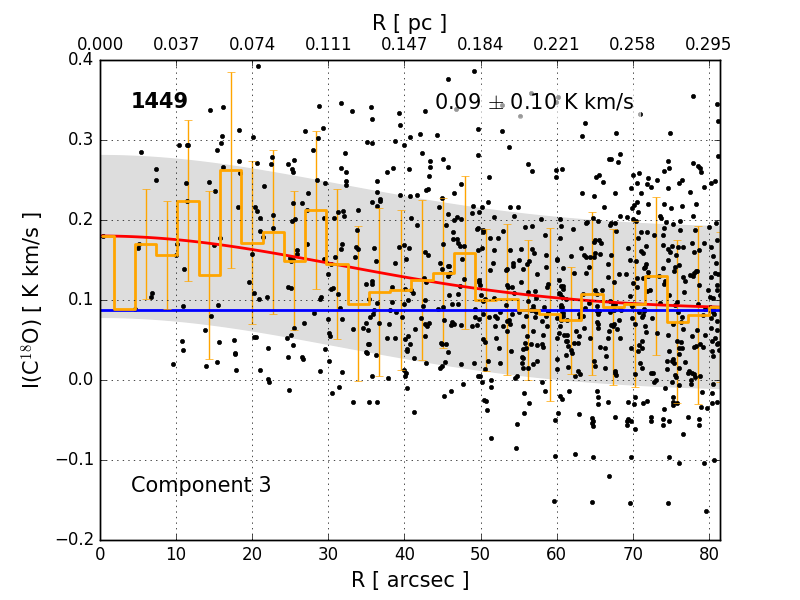}\\
      \caption{Radial profiles in CO (top row), \tCO\, (middle row), and \CeO\, (bottom row) integrated 
      intensities $W$ of the three components (columns) in s1449. Each pixel within a distance of $80\arcsec$
      is shown by a black dot. The running median is shown in orange, with the errorbars showing the standard
      deviation of $W$ in each radius bin. The Gaussian fit to the running median is shown in red, and the 
      offset in blue. The grey area shows the 1-sigma uncertainty on the best-fit profiles. The value of the 
      Gaussian amplitude and its uncertainty are shown in the top of each frame.}
      \label{fig:comp_radial_profile}
   \end{figure*}

   The radial profiles in integrated intensities $W$ of \CO, \tCO, and \CeO\, were 
   computed with a running median, and fit by a Gaussian function added to an
   offset in $W$. This offset is a measure of the background level. The Gaussian 
   parameters are the maximum integrated intensities $W_{\rm max}$, and the  
   velocity dispersion $\sigma$. We kept in the analysis of this paper only 
   sources with a significant $W_{\rm max}$, that is $W_{\rm max} / {\rm err}(W_{\rm max}) > 3$ 
   where ${\rm err}(W_{\rm max})$ is the fitting error on $W_{\rm max}$. This 
   enabled us to exclude the components corresponding to extended structures. As
   an example, Fig.~\ref{fig:comp_radial_profile} shows the radial profiles of
   the three components and three CO isotopologues of s1449. Only the second 
   component corresponds to a significantly peaked profile and is kept in the 
   analysis.


   \subsection{Temperature and density calculations}
   \label{sec:TexColdens}

   We calculated the excitation temperatures \Tex~ and column densities of \tCO~
   and \CeO~ following the method described by \citet{wilson_tools_2013}. The 
   radiative transfer equation, expressed with the main beam brightness 
   temperature \Tmb, gives:
   \begin{equation}
      \label{eq:rt1}
      \Tmb(\nu) = T_0 \left( \frac{1}{e^{T_0/T_{ex}}-1} 
                           - \frac{1}{e^{T_0/2.725} -1} \right)
                           (1 - e^{-\tau_{\nu}})
   \end{equation}
   where $T_0 = h\nu/k_{\rm B}$ (=5.53, 5.26 and 5.25 K for \CO, \tCO~ and \CeO,
   respectively), $\tau_{\nu}$ is the optical depth at the frequency $\nu$, and
   the beam filling factor was taken as 1 because the sources are extended clouds.

   In the case where the \CO\,(J=1-0) line is optically thick, we derive its
   excitation temperature:
   \begin{equation}
      \label{eq:Tex}
      \Tex = 5.53 / \ln \left[ 1 + \left( \frac{5.53}{\Tmb(\rm{CO}) + 0.836} \right) \right].
   \end{equation}
   In practice, we computed \Tex~ from the brightest channel of \Tmb(\rm{CO}).
   Since \CO~ was observed with FTS200 in parallel to the VESPA observations of
   \tCO~ and \CeO, it has a coarse resolution of $0.6$ km\,s$^{-1}$ and therefore an
   excellent SNR ($\gtrsim 50$). Hence, no further smoothing or Gaussian fitting
   was necessary to derive \Tex.

   Assuming that \CO, \tCO~ and \CeO~ have the same \Tex~ for the J=1-0 transition
   and that it is uniform along the sightline, the optical depth $\tau_{\nu}$ of
   each of \tCO~ and \CeO~ J=1-0 lines can be computed by inverting Eq.~(\ref{eq:rt1}).

   In our data, many pixels exhibit complex line profiles which cannot be fitted
   by a single Gaussian function. Therefore, instead of classically deriving the
   column density from the integrated line intensity as, for example, in \citet
   {feher_co_2017}, we directly integrate $\tau_{\nu}$:
   \begin{equation}
      \label{eq:Nco}
      N = Z(\Tex) \times 31.2 \times \frac{\nu_{10}^3}{A_{10}} \, \frac{\int \tau_\nu \, dv}{1 - \exp(-T_0/\Tex)}
   \end{equation}
   where we computed the partition function $Z(\Tex)$ using the Euler-Maclaurin 
   formula\footnote{The expansion to the third order is accurate to better than
   0.1\% for $\Tex>3$ K, and to better than $10^{-6}$ for $\Tex>12$ K.}, the
   Einstein coefficients were taken from the CDMS database\footnote
   {http://www.astro.uni-koeln.de/cdms/} \citep[$A_{10}($\tCO$)=6.33\times 10^{-8}$ s$^{-1}$, 
   $A_{10}($\CeO$)=6.27\times 10^{-8}$ s$^{-1}$]{endres_cologne_2016}, 
   $\nu_{10}$ is the rest frequency of the J=1-0 transition expressed in GHz, 
   and the integral is over the linewidth in $\mathrm{km\,s^{-1}}$.

   \subsection{Virial mass calculation}\label{sec:virial_analysis}

   To estimate the boundedness of sources, we computed the virial mass 
   $M_{\rm vir}$ of each core mapped with IRAM using the equation by \citet
   {maclaren_corrections_1988}:
   \begin{equation}
      M_{\rm vir} = k \, R \, \times 8 \ln 2 \, \sigma_{\rm H_2}^2,
   \end{equation}
   where $R$ is the radius of the core, $\sigma_{\rm H_2}$ is the total velocity
   dispersion of H$_2$, and $k=168$ $\mathrm {M_{\odot}\,pc^{-1}\,(km\,s^{-1})^{-2}}$ is
   a factor which assumes a Gaussian velocity distribution and that the density
   profile varies as $\rho \propto R^{-1.5}$ \citep{maclaren_corrections_1988}.

   In practice we computed the effective radius of the cores as the deconvolved
   geometric average of the semi-major $A$ and semi-minor $B$ axes provided in 
   the GCC catalogue for the opacity map: $R = \sqrt{AB-\theta^2}$, with
   $\theta=38.5\arcsec$ \citep{montillaud_galactic_2015}.
   The H$_2$ velocity dispersion $\sigma_{\rm H_2}$ was derived from the FWHM
   linewidth of a CO isotopologue:
   \begin{equation}
      \sigma_{\rm H_2} = \sqrt{ \frac{\kboltz T_{\rm kin}}{m_{\rm H_2}} 
      + {\sigma_{\rm NT}}^2}
   \end{equation}
   with ${\sigma_{\rm NT}}$ the non-thermal velocity dispersion defined as:
   \begin{equation}
      \sigma_{\rm NT} = \sqrt{ \frac{\Delta v^2}{8 \ln 2} - \frac{\kboltz T_{\rm kin}}{m} }
   \end{equation}
   where $\kboltz$ is the Boltzmann constant, $T_{\rm kin}$ is the kinetic 
   temperature, and $m$ and $\Delta v$ are the mass and the FWHM linewidth of 
   the CO isotopologue used for the calculation. The latter value was taken for 
   each component of each compact source from the multi-component Gaussian fits 
   presented in Sect.~\ref{sec:fit_examples}. The calculation was performed
   for both the \tCO\, and the \CeO\, lines to enable the comparison.

   Assuming that local thermal equilibrium (LTE) is satisfied, the kinetic
   temperature is equal to the excitation temperature \Tex. As in 
   Sect.~\ref{sec:TexColdens}, we used the optically thick \CO\, emission to
   estimate \Tex, except here the value of $\Tmb(\rm{CO})$ used in Eq.(~\ref{eq:Tex})
   is the peak value of the Gaussian fit to the analysed velocity component. As
   shown by \citet{feher_co_2017}, the linewidth fitting uncertainty results in 
   virial mass errors of less than 1\%, while optical depth effects in \tCO, 
   considering $\tau_{13}\lesssim3$, can lead one to overestimate the true FWHM
   by up to 30\% and therefore the virial mass by up to a factor of two. In
   Sect.~\ref{sec:results_source_class}, we present the virial analysis for both
   \tCO\, and \CeO, whose low optical depth alleviates the overestimation problem
   of \tCO. The distance uncertainty also affects the virial mass linearly, but
   since the present study focuses on a single cloud, it acts as a systematic 
   error rather than a random error on stability analysis.

   \subsection{Abundances}\label{sec:abundances}
   To estimate the gas mass from CO isotopologue observations, we assumed a 
   ratio $X_{\mathrm{CO/H_2}}=1.0 \times 10^{-4}$ \citep[e.g. ][]
   {pineda_relation_2010} and we used the Galactic gradient in C isotopes 
   reported by \citet{wilson_abundances_1994} to compute the ratios $X_{\CO/\tCO}
   = 75.9$ and $X_{\CO/\CeO}=618$ for the galactocentric distance of G202.3+2.5 
   \citep[9.11 kpc, ][]{montillaud_galactic_2015}.

\section{Source characterisation from IR data}

\subsection{Spectral energy distributions of GCC sources}

\begin{figure*}
   \includegraphics[width=\textwidth]{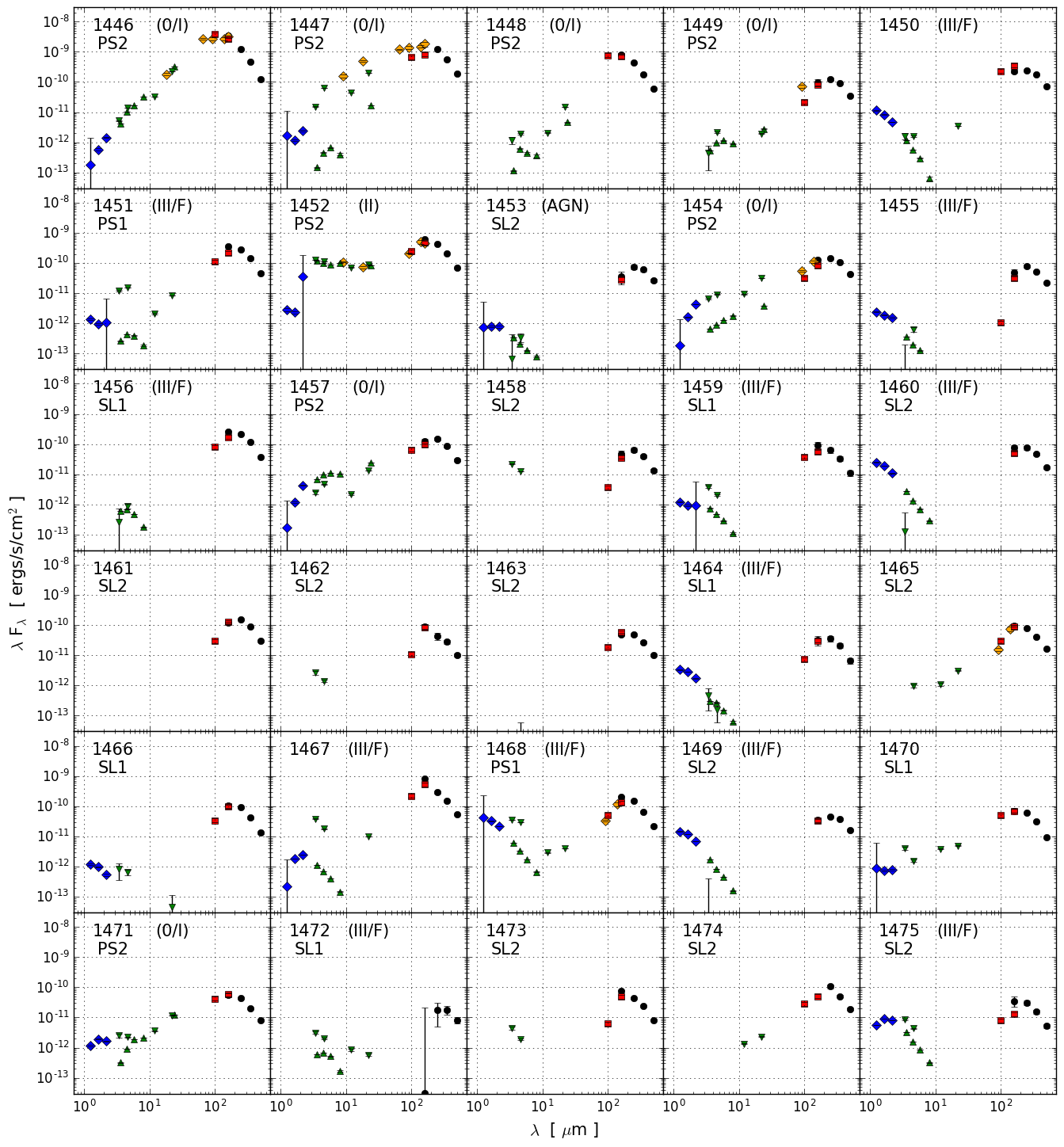}
   \caption{Spectral energy distributions of GCC sources 1446 to 1475. Blue 
   diamonds are 2MASS J, H, and Ks fluxes from the 2MASS PSC. Green tip-up 
   triangles are {\it Spitzer}
   3.6, 4.5, 5.8, 8.0, and 24\,$\mu$m fluxes from the YSO catalogue of \citet
   {rapson_spitzer_2014}. Green tip-down triangles are WISE 3.4, 4.6, 12, and 22
   \,$\mu$m fluxes from our photometry measurements. Orange diamonds are AKARI
   9, 18, 65, 90, 140, and 160\,$\mu$m fluxes from the AKARI MIR and FIR point
   source catalogues. Red squares show the fluxes of PACS 100 and 160\,$\mu$m
   from our photometry measurements. Black circles are fluxes of PACS 160 and
   SPIRE 250, 350, and 500\,$\mu$m from the GCC catalogue of compact sources.
   For the fluxes selected from public point source catalogues, only the source 
   nearest to the GCC source centre was considered. Error bars show the flux
   uncertainties, but can be smaller than the marker size. The source number in
   the GCC catalogue is indicated in the top-left corner of each frame, along
   with our evolutionary state classification (SL and PS stand for starless and
   protostellar, respectively, and the numbers 1 and 2 for candidate and
   robust classification, respectively). The Spitzer-based classification by 
   \citet{rapson_spitzer_2014} in shown in parentheses. }
   \label{fig:allSEDs1}
\end{figure*}

\begin{figure*}
   \includegraphics[width=\textwidth]{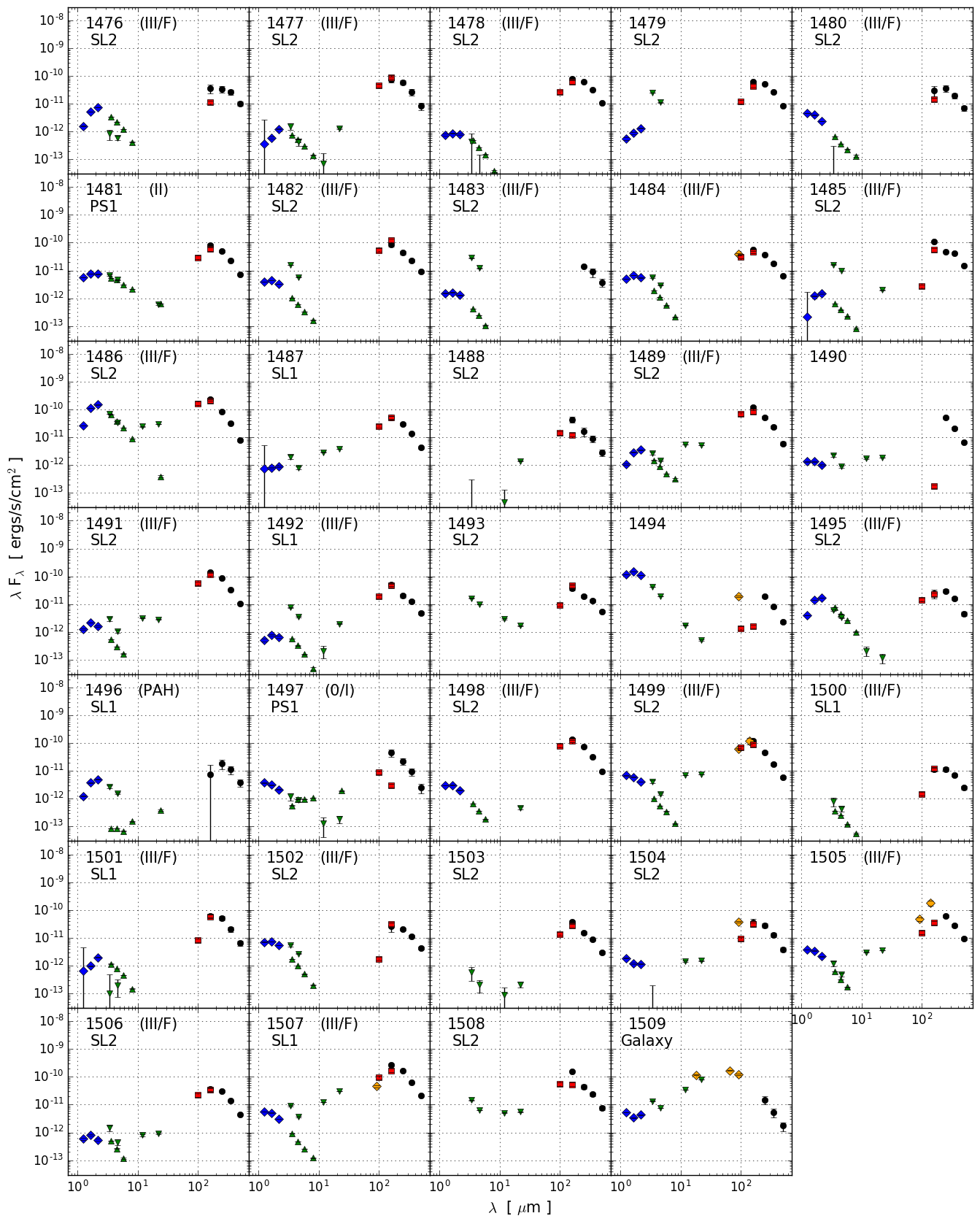}
   \caption{Same as Fig.~\ref{fig:allSEDs1} for the GCC sources 1476 to 1509.}
   \label{fig:allSEDs2}
\end{figure*}

\begin{figure*}
   \includegraphics[width=\textwidth]{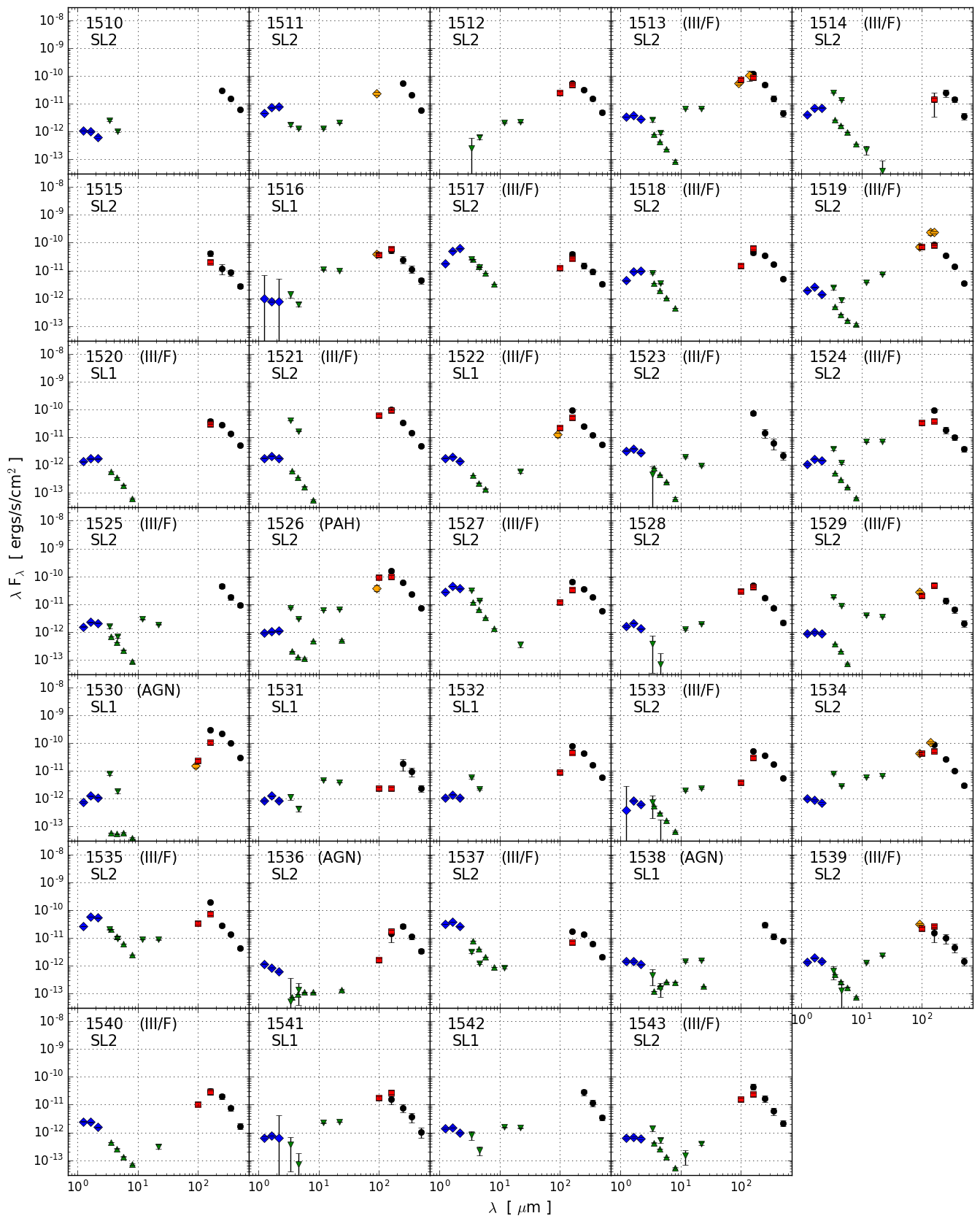}
   \caption{Same as Fig.~\ref{fig:allSEDs1} for the GCC sources 1510 to 1543.}
   \label{fig:allSEDs3}
\end{figure*}

   To classify and characterise each source, we considered its spectral energy
   distribution from NIR to FIR wavelengths, as presented in Figs.~\ref
   {fig:allSEDs1}-\ref{fig:allSEDs3}. We used the total flux estimates 
   and their associated uncertainties provided for PACS 160\,$\mu$m and the 
   three SPIRE bands in the GCC catalogue. In addition, we computed the total
   fluxes and their associated uncertainties for PACS 100\,$\mu$m, and the four
   WISE bands at 3.4, 4.6, 12 and 22\,$\mu$m directly from their maps. To do so,
   we summed the emission of pixels near the source weighted by a 2D-Gaussian
   defined with the ellipse properties reported for the PACS 160\,$\mu$m in the
   GCC catalogue ('AFWHM1', 'BFWHM1', and 'PA1'), and cut at a distance of $3\sigma$
   from the source centre. We subtracted the background estimated from the median
   of the pixels in an elliptical annulus between $3.5\sigma$ and $5\sigma$ from
   the source centre. The same method applied to the PACS 160\,$\mu$m map provides
   results in good agreement with the values in the GCC catalogue, with approximately 55\% 
   of sources with relative differences lesser than 30\%, and approximately 40\% between 
   30\% and 100\%. Disagreement typically occurs for sources presenting a complex
   background like gradients or nearby bright compact sources (as for
   example in s1488, Fig.~\ref{fig:Source_analysis_1488}).

   We complemented these values with the fluxes found in the 2MASS point source
   catalogue (PSC, bands J, H, and K$_s$), the AKARI mid- and far-IR PSCs (bands
   at 9, 18, 65, 90, 140, and 160\,$\mu$m), as well as the \textit{Spitzer} 
   sources reported in the YSO catalogue by \citet{rapson_spitzer_2014}, at 3.6,
   4.5, 5.8, 8.0, and 24\,$\mu$m. For each GCC source, only the nearest source
   from each of the 2MASS, AKARI, and {\it Spitzer} catalogues was considered.
   In some cases several point sources could be associated with the GCC source,
   but a single candidate YSO is sufficient to make the GCC source a candidate
   protostellar core. The YSO catalogue by \citet{rapson_spitzer_2014} was
   compiled from cuts in colour-magnitude and colour-colour diagrams. Finally, 
   we also correlated the GCC sources with the YSO catalogue by \citet
   {marton_all-sky_2016}. This 
   catalogue was built using a support vector machine method on the WISE PSC, 
   leading to an all-sky YSO catalogue. The cross-correlation between this 
   catalogue and the GCC catalogue provides only a few matches, therefore we did
   not use the fluxes reported in this catalogue.

\subsection{Starless vs. protostellar: Classification scheme}
   \label{sec:GCCsources_method}

   From the analysis of the SEDs, a first phase of classification of GCC sources
   was performed as follows. Sources without any associated emission at 
   wavelengths shorter than 100\,$\mu$m were classified as reliable starless 
   cores (code 2 in our tables). Sources with an associated mid-IR source 
   classified as class 0/I or class II by \citet{rapson_spitzer_2014} were 
   classified as protostellar.
   In most cases, additional evidence that the core is protostellar could be 
   found, like the outflows and Herbig-Haro objects associated with s1446 and s1447, 
   or the fact that the compact source is visible at all wavelengths
   from 3 to 500 $\mu$m. The source was then considered a reliable protostellar
   core (code 4 in our tables). For s1481 and s1497, no compact emission was 
   detected at 100 $\mu$m and the sources were classified as candidate protostellar 
   cores (code 3 in our tables). Rapson et al. classification of active galactic
   nuclei (AGNs) and polycyclic aromatic hydrocarbon (PAH)
   emission was also used to identify six starless cores. 

   In the second phase, for the remaining sources, the morphology of the emission
   in all WISE and \textit{Herschel} maps were checked by eye to identify whether
   it corresponds to a point (or compact) source associated with the core, or to
   structured background emission. In the latter case, the source was classified
   as a candidate starless core (code 1 in our tables) if a 2MASS point source
   was detected, and as a reliable starless core if no 2MASS emission was found. 
   When mid-IR compact emission was observed, we searched for evidence that it 
   is (or is not) physically associated with the GCC source. Sources where a 
   large spatial shift ($>80\%$ of the half width of the GCC source) between the
   compact mid-IR and far-IR 
   emissions is observed (e.g. s1479) were considered reliable starless cores.
   We also searched the Gaia DR2 archive \citep{gaia_collaboration_gaia_2016, 
   venuti_gaia-eso_2018} for an optical counterpart of the near- and mid-IR source
   to compare its parallax with the cloud distance (1.3 mas for a distance of
   760 pc). This enabled us to rule out about 40 IR sources whose
   parallaxes are incompatible with that of the cloud. The corresponding GCC
   sources (e.g. s1460) were classified as candidate starless cores if $1.3 
   {\rm mas} \not\in [\pi-\delta\pi, \pi+\delta\pi]$, where $\pi$ is the source
   parallax and $\delta\pi$ its uncertainty, and as reliable starless
   cores if $1.3{\rm mas} \not\in [\pi-3\delta\pi, \pi+3\delta\pi]$. In two cases
   (s1451 and s1468), compact emission is visible at all 
   wavelengths from 3 to 500 $\mu$m, and we classified the sources as candidate
   protostellar sources. In the case of s1468, the parallax is compatible with
   that of the cloud. Those two sources were classified as 'class III/field star'
   by \citet{rapson_spitzer_2014}, and might be some of the few genuine class III
   in this category which is very likely to be dominated by field stars. For this
   reason, except for s1451 and s1468, we classified as candidate starless those
   source where the mid-IR emission was classified as class III/field star by
   \citet{rapson_spitzer_2014}. Finally, the sources with column density lower
   than $2 \times 10^{21}\,{\rm cm}^{-2}$ were classified as candidate starless
   cores, since this range of densities corresponds to low-density translucent
   cores rather than dense cores \citep[e.g.][]{zhao_mercapto_2015}. When cores
   were classified as candidate starless cores from several independent criteria,
   we upgraded them to reliable starless cores.

   This scheme enabled us to propose a classification for the 98 GCC sources as
   summarised in Table~\ref{tab:stateclass}. However, it failed to identify a 
   few special cases. We changed the classification of s1509 to extragalactic
   because of its SED which peaks at 50 $\mu$m, its very low column density ($8.4
   \times 10^{20}\,{\rm cm}^{-2}$) and its location very isolated from the 
   filaments. For seven sources, we chose to keep an 'undetermined' classification
   because of contradictory arguments, as summarised in the last column of 
   Table~\ref{tab:stateclass}. This is for example the case of s1450, the densest
   and most massive source of this region, where the Gaia, 2MASS, and mid-IR 
   detections correspond to a point source that is unlikely to be physically
   related with the core, but where the compact and relatively strong emission
   at 100 $\mu$m is compatible with a class 0 object. The lack of data near 70\,$\mu$m
   makes it difficult to conclude.

   In addition, we examined the dust temperature profiles of each source looking
   for significant radial increase or decrease in \Tdust. From the \Tdust\, map,
   the pixels within a radius $R_{\rm max}=1.5 \times A_{\rm FWHM}$, where 
   $A_{\rm FWHM}$ is the source major axis FWHM, were plotted as a function of 
   the distance to the source centre, and the profile was fitted by a 1D-Gaussian 
   function. The uncertainty $\delta T$ on the amplitude $\Delta T$ was computed
   as the standard deviation of the residuals for $r<0.5 R_{\rm max}$. An increase 
   or decrease in \Tdust\, was considered significant where $|\Delta T| / \delta T>3$.
   Table~\ref{tab:stateclass} shows that s1446 is the only protostellar core
   showing a significant increase in temperature towards the centre of the core. 
   For the other protostellar sources, the temperature profile often decreases
   towards the centre of the core, probably because of the low physical resolution
   of our \Tdust\, maps at this distance ($40\arcsec$ corresponding to 0.15 pc at
   760 pc). Interestingly, the other five sources with a significant \Tdust\,increase
   towards their centre (s1489, s1499, s1519, s1526, and s1531) have column densities
   of the order of $2\times 10^{21}\,{\rm cm}^{-2}$ and relatively warm 
   temperatures ($\gtrsim 15$ K). Most of them also show bright PAH emission 
   on their edges or in their volume. Several other translucent cores show similar
   mid-IR morphologies, and were tagged as 'warm starless' in the last column
   of Table~\ref{tab:stateclass}. We conclude that the \Tdust\,profiles are not
   a good indication of the starless or protostellar nature of the cores at this
   distance (and with this resolution), and we do not use them in our classification
   scheme.

   The results of this classification are analysed in Sect.~\ref{sec:results_source_class}.

\section{Summary plots of noteworthy sources}

\begin{figure*}
   \includegraphics[width=\textwidth]{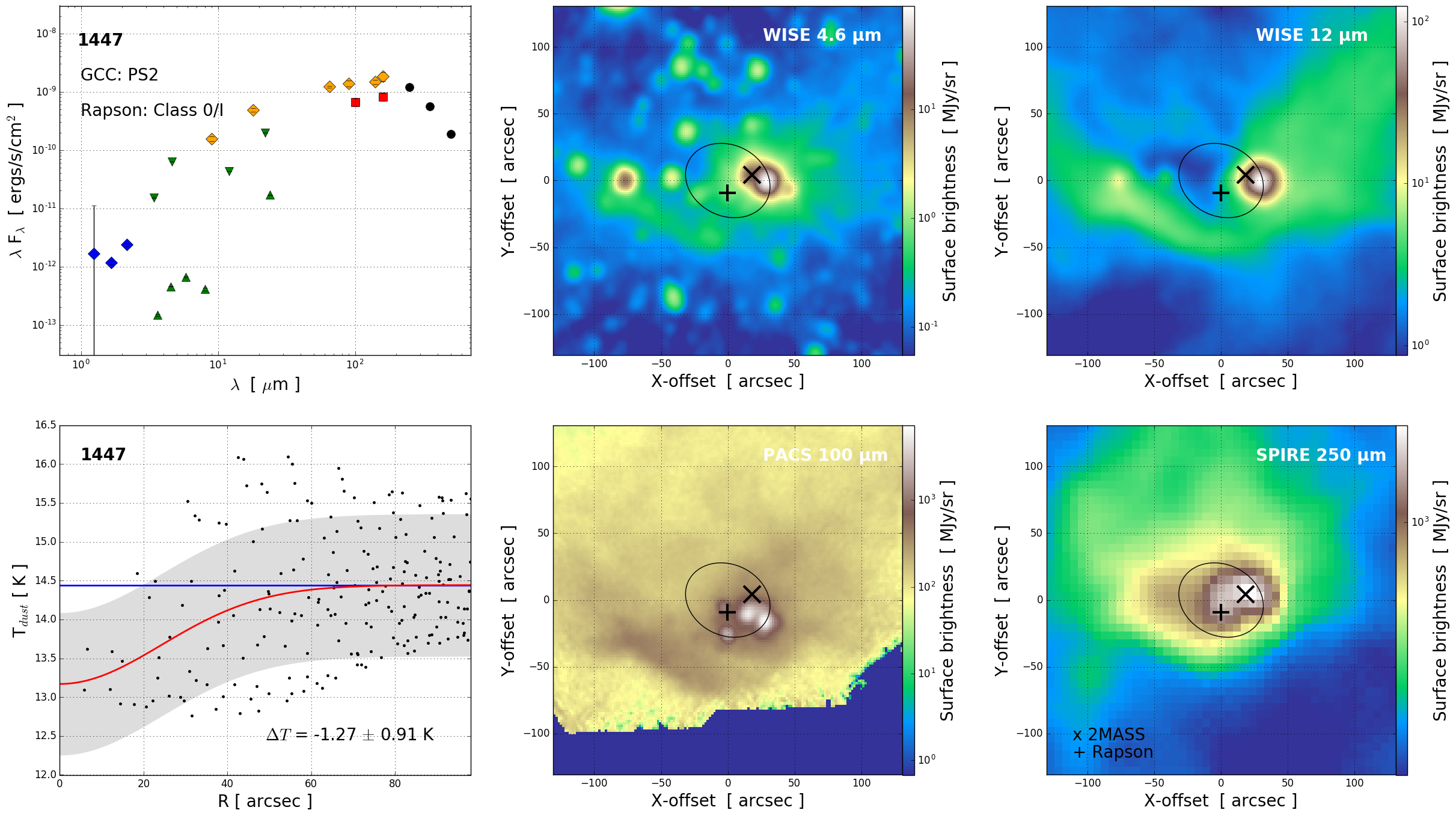}
   \caption{Same as Fig.~\ref{fig:source_class_example} for the GCC source 1447, 
   also referred to as IRAS 25 or NGC 2264 O. The cross and plus symbols
   show the locations of the 2MASS and Spitzer \citep[as reported by][]
   {rapson_spitzer_2014} point sources, respectively.}
   \label{fig:Source_analysis_1447}
\end{figure*}

\begin{figure*}
   \includegraphics[width=\textwidth]{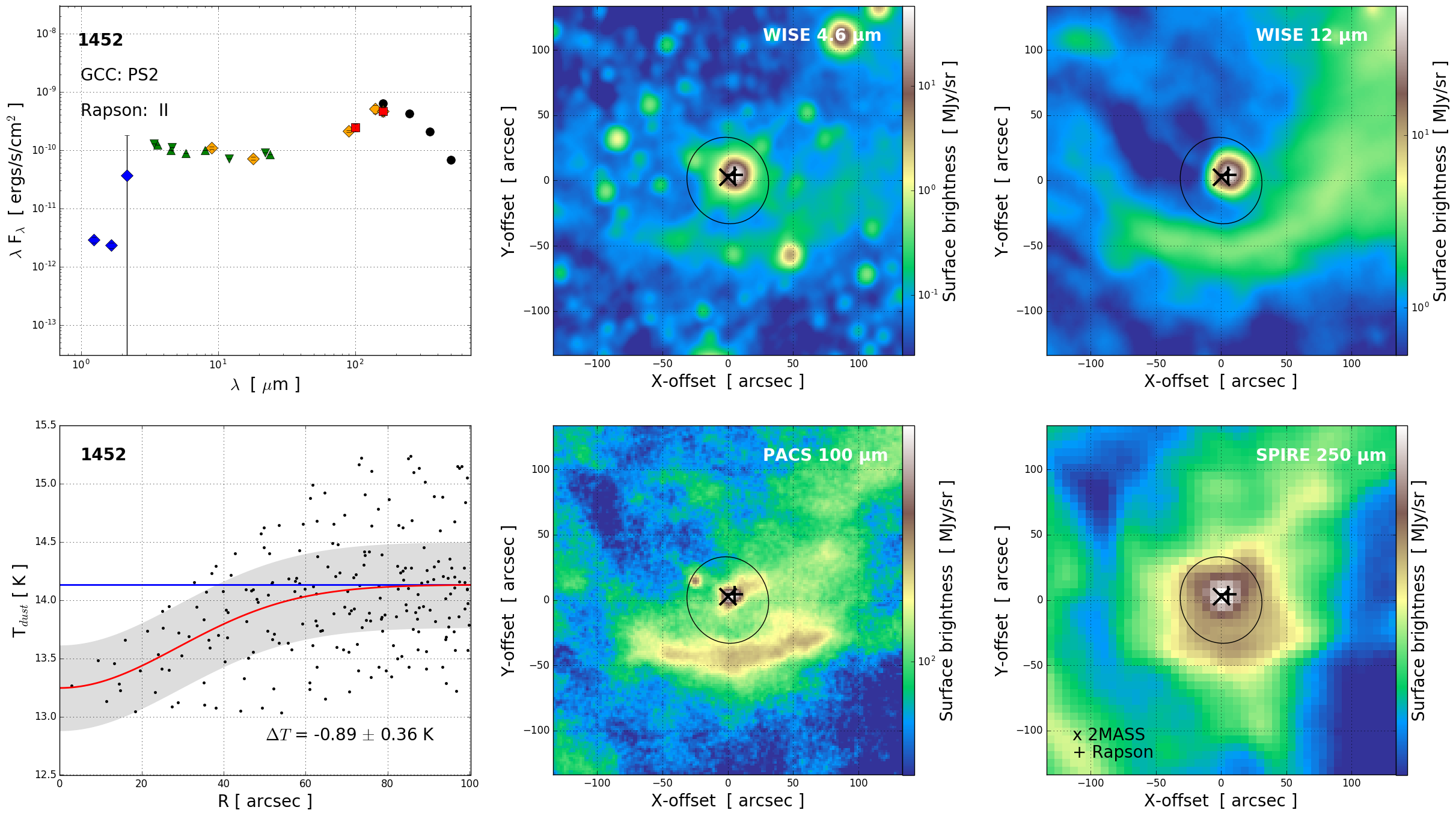}
   \caption{Same as Fig.~\ref{fig:Source_analysis_1447} for the GCC source 1452.}
   \label{fig:Source_analysis_1452}
\end{figure*}

\begin{figure*}
   \includegraphics[width=\textwidth]{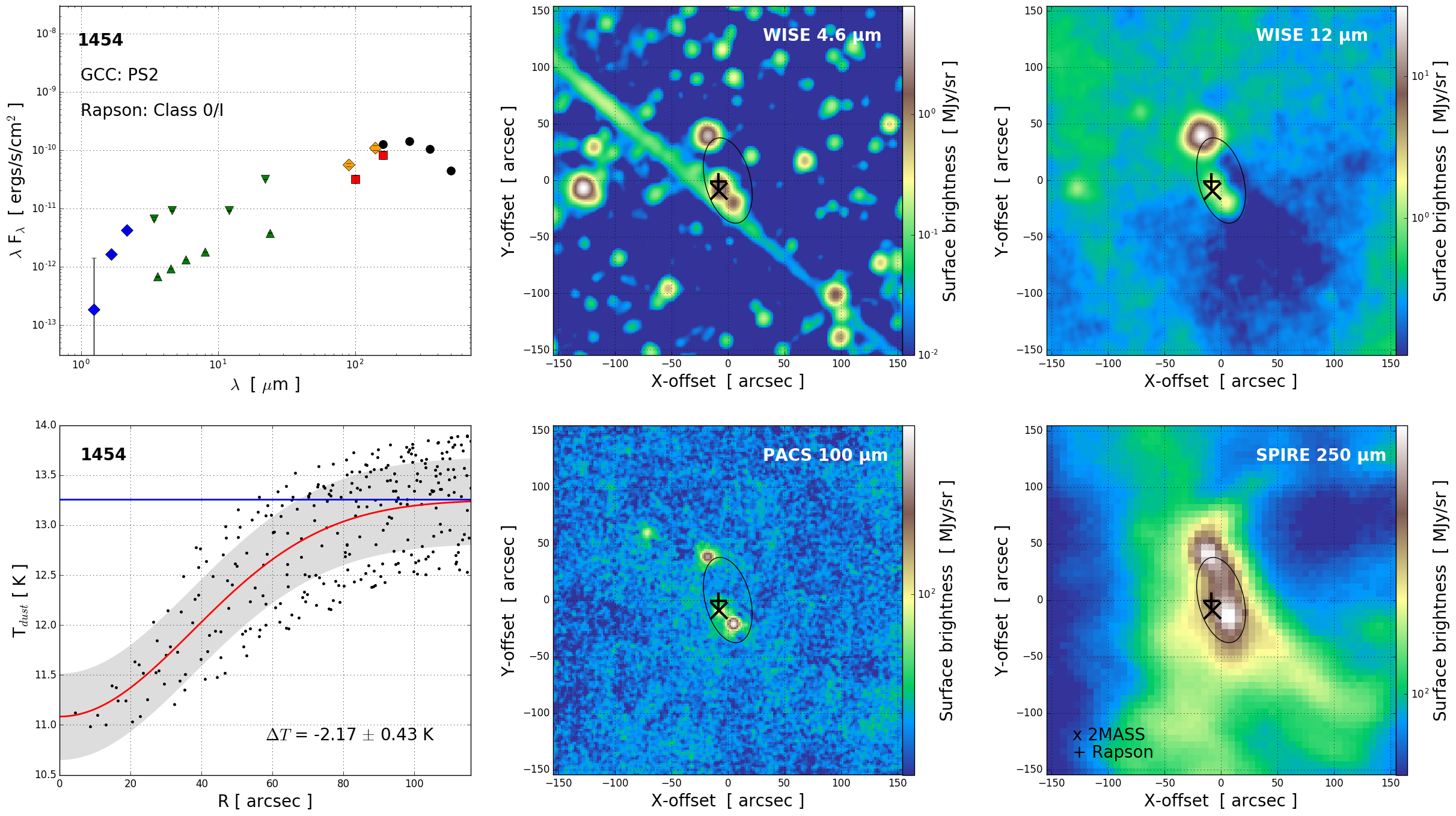}
   \caption{Same as Fig.~\ref{fig:Source_analysis_1447} for the GCC source 1454, 
   also referred to as the north clump in this paper. The straight line in the 
   WISE 4.6\,$\mu$m map is an artefact caused by a bright star $\sim 7\arcmin$
   in the north-east direction.
   }
   \label{fig:Source_analysis_1454}
\end{figure*}

\begin{figure*}
   \includegraphics[width=\textwidth]{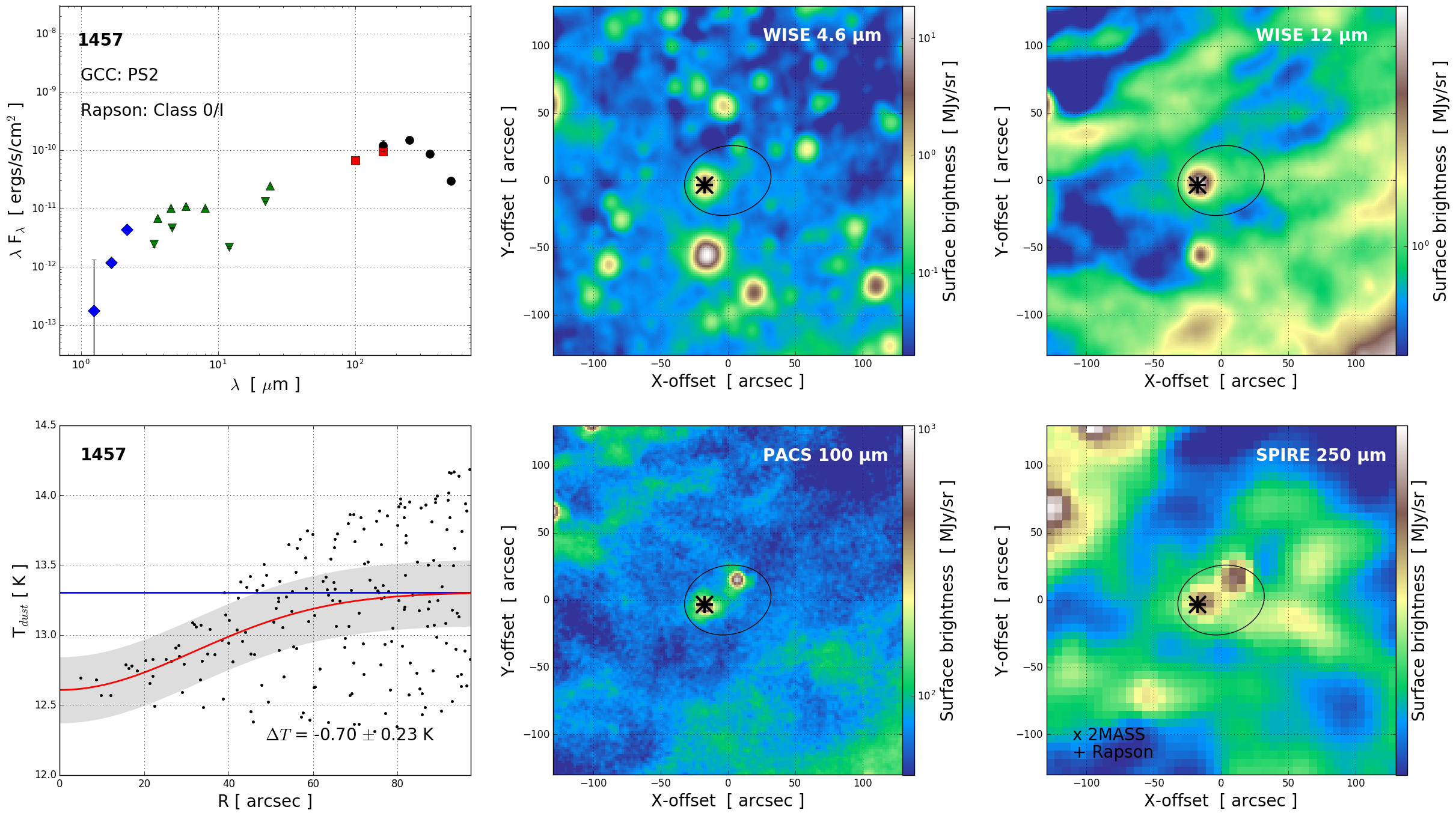}
   \caption{Same as Fig.~\ref{fig:Source_analysis_1447} for the GCC source 1457.}
   \label{fig:Source_analysis_1457}
\end{figure*}

\begin{figure*}
   \includegraphics[width=\textwidth]{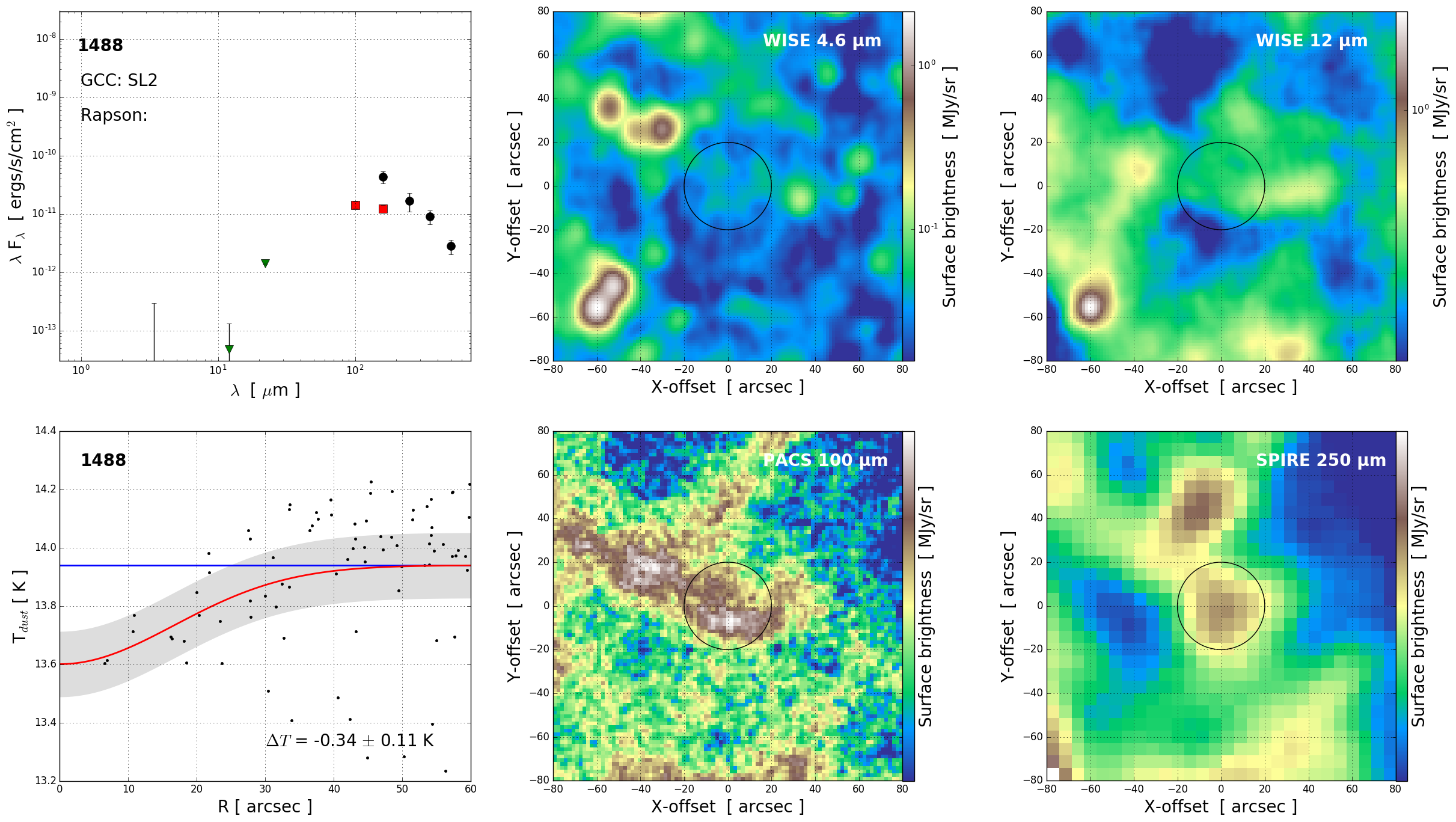}
   \caption{Same as Fig.~\ref{fig:Source_analysis_1447} for the GCC source 1488.}
   \label{fig:Source_analysis_1488}
\end{figure*}

\end{document}